\documentclass[11pt,nofootinbib,showpacs]{revtex4}
\usepackage{amsfonts,amssymb,amsmath,graphicx,multirow,dsfont}
\def\dac{\displaystyle\frac}

\def\[{\left[}
\def\]{\right]}
\def\({\left(}
\def\){\right)}

\def\ot{\leftarrow}

\newcommand{\diag}{\mathop{\rm diag}\nolimits}
\newcommand{\const}{\mathop{\rm const}\nolimits}
\begin{document}

\baselineskip7mm

\title{Dynamics of the cosmological models with perfect fluid in Einstein-Gauss-Bonnet gravity: low-dimensional case}

\author{Sergey A. Pavluchenko}
\affiliation{Programa de P\'os-Gradua\c{c}\~ao em F\'isica, Universidade Federal do Maranh\~ao (UFMA), 65085-580, S\~ao Lu\'is, Maranh\~ao, Brazil}

\begin{abstract}
In this paper we performed investigation of the spatially-flat cosmological models whose spatial section is product of three- (``our Universe'') and extra-dimensional parts.
The matter source  chosen to be the perfect fluid which exists in the entire space.
We described all physically sensible cases for the entire range of possible initial conditions and parameters as well as
brought the connections with vacuum and $\Lambda$-term regimes described earlier.
In the present paper we limit ourselves with $D=1, 2$ (number
of extra dimensions).
The results suggest that in $D=1$ there are no realistic compactification regimes while in $D=2$ there is if $\alpha > 0$ (the Gauss-Bonnet coupling) and
the equation of state $\omega < 1/3$; the
measure of the initial conditions leading to this regime is increasing with growth of $\omega$ and reaches its maximum at $\omega \to 1/3 - 0$.
We also describe some pecularities of the model, distinct to the vacuum and $\Lambda$-term cases -- existence of the isotropic power-law regime,
different role of the constant-volume solution and the presence of the maximal density
for $D = 2$, $\alpha < 0$ subcase and associated features.
\end{abstract}

\pacs{04.50.-h, 11.25.Mj, 98.80.Cq}






\maketitle

\section{Introduction}

The idea of the extra spatial dimensions is quite old -- it predecesses even General Relativity (GR). Indeed, the first ever multidimensional model was proposed by
Nordstr\"om in 1914~\cite{Nord1914} and it was built on Nordstr\"om's second gravity theory~\cite{Nord_2grav} and Maxwell's electromagnetism. Though, as we know, the
Nordstr\"om's scalar gravity was proved to be wrong and soon almost forgotten, yet, his idea of extra dimensions survived and transformed into what is now called
Kaluza-Klein theory~\cite{KK1, KK2, KK3}. It is interesting to note that Kaluza-Klein theory unified all interaction which were known at that time. By now we know more
fundamental interactions and so to unify them all in the same manner we will need more spatial dimensions. Nowadays, the promising candidate for unified theory is
string/M theory.

When people started to study gravitational counterpart of string theories, they discovered terms in the Lagrangian which are quadratic with respect to the curvature;
particular examples include $R^2$ and
$R_{\mu \nu} R^{\mu \nu}$ terms~\cite{sch-sch} in the Lagrangian of the Virasoro-Shapiro
model~\cite{VSh1, VSh2}, $R^{\mu \nu \lambda \rho}
R_{\mu \nu \lambda \rho}$ term~\cite{Candelas_etal} for the low-energy limit
of the $E_8 \times E_8$ heterotic superstring theory~\cite{Gross_etal} and others. An important milestone was a discovery that there exist only one combination of the above
terms which leads to ghost-free nontrivial gravitation interaction -- the Gauss-Bonnet (GB) term~\cite{zwiebach}

$$
L_{GB} = L_2 = R_{\mu \nu \lambda \rho} R^{\mu \nu \lambda \rho} - 4 R_{\mu \nu} R^{\mu \nu} + R^2.
$$

\noindent This term was already known at that time, as it was first discovered
by Lanczos~\cite{Lanczos1, Lanczos2} (and therefore sometimes it is referred to
as the Lanczos term). It is an Euler topological invariant in (3+1) and lower dimensions while in (4+1) and higher it gives nontrivial contribution to the equations of
motion. Soon after the idea of~\cite{zwiebach} was supported and extended in~\cite{zumino} on the case of higher-then-quadratic terms. So that the unified theory in the
final form should include contributions from all possible curvature powers. In this regard the Lovelock gravity~\cite{Lovelock}, which is the sum of all possible Euler
topological invariants, which give nontrivial
contribution to the equations of motion in particular number of dimensions, is of special interest. Of course, the Einstein-Gauss-Bonnet (EGB) gravity, which is under
consideration in the current paper, is a particular case of more general Lovelock gravity.

Our everyday experience, as well as experiments in particle and space physics, clearly indicate that there are only three spatial dimensions (for example, in Newtonian gravity
in more then three spatial dimensions there are no stable orbits while we are rotating around the Sun for ages). So that to bring together the extra-dimensional theories and
the experiment, we need to explain where are these extra dimensions. The commonly accepted answer is that the extra dimensions are compactified on a very small scale -- similar
to the ``curling'' of extra dimension into a circle in the original Kaluza-Klein paper. But this answer, in turn, gives rise to another one -- how come that they became compact?
The answer to this question is not so simple. The first attempts to answer this question involves solution known as ``spontaneous compactification''~\cite{add_1, Deruelle2}.
Similar solutions but more relevant to cosmology were proposed in~\cite{add_4, Deruelle2} (see also~\cite{prd09}). More natural way to achieve compactified extra dimensions
is ``dynamical compactification''. The works on this direction involves different approaches~\cite{add_8, add13} and different setups~\cite{MO04, MO14}. Also, apart from the
cosmology, the studies of extra dimensions involve investigation of
properties of black holes in Gauss-Bonnet~\cite{alpha_12, add_rec_1, add_rec_2, add_o_1, add_o_2, add_o_3, addn_1, addn_2} and
Lovelock~\cite{add_rec_3, add_rec_4, addn_3, addn_4, addn_4.1} gravities, features of gravitational collapse in these
theories~\cite{addn_5, addn_6, addn_7}, general features of spherical-symmetric solutions~\cite{addn_8}, and many others.

As we shall see, the equations of motion for EGB and for more general Lovelock gravity are quite complicated and it is nontrivial to find exact solutions within them,
so that one usually apply {\it ansatz} of some sort to obtain them. For cosmology the usual {\it ans\"atzen} are power-law and exponential -- the former resemble Friedmann
stage while the latter -- accelerated expansion nowadays or inflationary stage in Early Universe. Power-law solution were studied in~\cite{Deruelle1, Deruelle2} and more
 recently in~\cite{mpla09, prd09, Ivashchuk, prd10, grg10}, so that we can say that we have some understanding of their dynamics. One of the first considerations of the
 exponential solutions in the considered theories was done in~\cite{Is86}, the recent works include~\cite{KPT}, the separate description of the cases with variable~\cite{CPT1}
 and constant~\cite{CST2} volume; see also~\cite{PT} for the discussion of the link between existence of power-law and exponential solutions as well as for the discussion
 about the physical branches of the  solutions.
 We have also described the general scheme for building all possible exponential solutions in arbitrary dimensions and with arbitrary
 Lovelock contributions taken into account~\cite{CPT3}. Deeper investigation revealed that not all of the solutions found in~\cite{CPT3} could be called ``stable''~\cite{my15};
 see also~\cite{iv16} for more general approach to the stability of exponential solutions in EGB gravity.

 The above approach gave us asymptotic power-law and exponential solutions, but not the evolution. Without full evolution we cannot decide if the asymptotic solution which we
 found realistic or not, could it be reached from vast enough initial conditions or is it just some artifact. To answer this question we need to go beyond exponential or
 power-law {\it ans\"atzen} and consider generic form of the scale factor. In this case, though, one would often resort to numerics. In~\cite{CGP1} we considered the
 cosmological model with the spatial part being a product of three- and extra-dimensional spatially curved manifolds and numerically demonstrated the existence of
 the phenomenologically
sensible regime when the curvature of the extra dimensions is negative and the EGB theory does not admit a maximally symmetric solution. In this case both the
three-dimensional Hubble parameter and the extra-dimensional scale factor asymptotically tend to the constant values (so that three-dimensional part expanding exponentially while
the extra dimensions tends to a constant ``size'').
In~\cite{CGP2} the study of this model was continued and it was demonstrated that the described above regime is the only realistic scenario in the considered model.
 Recent analysis of the same model~\cite{CGPT} revealed that, with an additional constraint on couplings, Friedmann-type late-time behavior
could be restored.

This investigation was performed numerically, but to find all possible regime for a particular model, the numerical methods are not the best practice. We have noted that if we
consider the model with spatial section as the product of three- and extra-dimensional spatially flat subspaces, the equations of motion simplify and become first-order instead
of the second one\footnote{The situation similar to the Friedmann equations -- if the spatial curvature $k \ne 0$ then they are second order with respect to the scale
factor $a(t)$ ($\ddot a$ is the highest derivative), but if $k \equiv 0$ then they could be rewritten in terms of Hubble parameter $H \equiv \dot a/a$ and become first order
($\dot H$ is the highest derivative).}. In this case the dynamics could be analytically described with use of phase portraits and so we performed this analysis. For vacuum EGB
case it was done in~\cite{my16a} and reanalyzed in~\cite{my18a}. The results suggest that in the vacuum model has two physically viable regimes -- first of them is the smooth transition from high-energy GB Kasner to low-energy GR Kasner. This regime
exists for $\alpha > 0$ (Gauss-Bonnet coupling) at $D=1,\,2$ (the number of extra dimensions) and for $\alpha < 0$ at $D \geqslant 2$ (so that at $D=2$ it appears for both signs of $\alpha$). Another viable regime is the smooth
transition from high-energy GB
Kasner to anisotropic exponential regime with expanding three-dimensional section (``our Universe'') and contracting extra dimensions; this regime occurs only for $\alpha > 0$ and at $D \geqslant 2$. Apart from the EGB, similar analysis but for cubic Lovelock contribution taken into account was performed in~\cite{my18b, my18c}.

The similar analysis for EGB model with $\Lambda$-term was performed in~\cite{my16b, my17a} and reanalyzed in~\cite{my18a}. The results suggest that the only realistic scenario
is the transition from high-energy GB Kasner to anisotropic exponential
solution, it requires $D \geqslant 2$, see~\cite{my16b, my17a, my18a} for exact limits on ($\alpha, \Lambda$).
The low-energy GR Kasner is
forbidden in the presence of the $\Lambda$-term so the corresponding transition do not occur.

In the studies described above we have made two important assumptions -- we considered both subspaces being isotropic and spatially flat.
But what will happens in we lift these conditions?
Indeed, the spatial section
being a product of two isotropic spatially-flat subspaces could hardly be called ``natural'', so that we considered the effects of anisotropy and spatial curvature in~\cite{PT2017}. The effect of the initial anisotropy could be demonstrated on the example of vacuum $(4+1)$-dimensional EGB cosmology with Bianchi-I-type metric
(all the directions are independent), where the only future asymptote is nonstandard singularity~\cite{prd10}. Our analysis~\cite{PT2017} suggest that the transition from
 GB Kasner
regime to anisotropic exponential expansion (with expanding
three and contracting extra dimensions) is stable with respect to breaking the symmetry within both three- and extra-dimensional subspaces. However, the details of the dynamics in
$D=2$ and $D \geqslant 3$ are different -- in the latter there exist anisotropic exponential solutions with ``wrong'' spatial splitting and all of them are accessible from generic
initial conditions. For instance, in $(6+1)$-dimensional space-time there are anisotropic exponential solutions with $[3+3]$ and $[4+2]$ spatial splittings, and some of the initial
conditions in the vicinity of $E_{3+3}$ actually end up in $E_{4+2}$ -- the exponential solution with four and two isotropic subspaces. In other words, generic initial conditions
could easily end up with ``wrong'' compactification, giving ``wrong'' number of expanding spatial dimensions (see~\cite{PT2017} for details).

The effect of the spatial curvature on the cosmological dynamics could be dramatic -- say, positive curvature changes the inflationary asymptotic~\cite{infl1, infl2}.
In EGB gravity the influence of the spatial curvature (also described in~\cite{PT2017})
reveal itself only if the curvature of the extra dimensions is negative and $D \geqslant 3$
-- in that case there exist  ``geometric frustration'' regime, described in~\cite{CGP1, CGP2} and further investigated in~\cite{CGPT}.

The current manuscript is, from one hand, a direct continuation of our ongoing study of all possible cosmologically sensible cases in EGB gravity:
vacuum models (see~\cite{my16a}) and
models with $\Lambda$-term (see~\cite{my16b, my17a}; both cases with revised presentation of the results are reviewed in~\cite{my18a}).
Both of them are important milestones on our way to understand the cosmological dynamics of EGB (and more general Lovelock) gravity, but if we want to describe
our Universe (and this is the goal of any realistic physical theory regardless of the field), then we must
make the theory as close as possible to the reality. And in the reality we observe not vacuum and not only $\Lambda$-term, but ordinary matter as well.
So that in current paper we started to
consider another important matter source for EGB gravity -- matter in form of perfect fluid -- making the study of the entire EGB case more complete.
It worth mentioning that we already
performed consideration of some particular GB and EGB models with matter source in form of perfect fluid: some initial considerations in~\cite{KPT}, some deeper studies of
(4+1)-dimensional
Bianchi-I case in~\cite{prd10} and deeper investigation of power-law regimes in pure GB gravity in~\cite{grg10}. In the last cited paper we obtained the results on which we
will rely in the current study
and which we are going to use to explain some of the features. The most important of these results is the fact that $\omega = 1/3$ is the critical value for the equation of
state which interchange the dominance of GB and matter terms: for $\omega < 1/3$ past asymptote is GB-dominated and future asymptote is matter-dominated, for $\omega > 1/3$ the
situation changes -- past asymptote is matter-dominated while future asymptote is GB-dominated. But there is one important issue which disallow us from directly implementing these
results to our current study -- now we have EGB gravity, so that the low-energy regime now is described by Einstein-Hilbert Lagrangian and for it the limiting equation of state is
$\omega = 1$ (Jacobs solution~\cite{Jacobs}). And since we do not consider $\omega > 1$, all of our low-energy regimes are the same so the difference lies only in high-energy ones. So that the analogue with GB case is not direct and there could be other interesting features and observations.

The manuscript is structured as follows: in the following Section~\ref{s1} we write down equations of motion, then consider the dynamics for $D=1$ in Section~\ref{d1} and for $D=2$
in Section~\ref{d2}, discuss the
obtained results and the interesting observations in Section~\ref{dis} and finally draw conclusions in Section~\ref{conc}.

\section{Equations of motion}
\label{s1}

The structure of the general Lovelock gravity~\cite{Lovelock} is the following: its Lagrangian consists of terms

\begin{equation}
L_n = \frac{1}{2^n}\delta^{i_1 i_2 \dots i_{2n}}_{j_1 j_2 \dots
j_{2n}} R^{j_1 j_2}_{i_1 i_2}
 \dots R^{j_{2n-1} j_{2n}}_{i_{2n-1} i_{2n}}, \label{lov_lagr}
\end{equation}

\noindent where $\delta^{i_1 i_2 \dots i_{2n}}_{j_1 j_2 \dots
j_{2n}}$ is the generalized Kronecker delta of the order $2n$.
One can note that $L_n$ is the Euler invariant in $D < 2n$ spatial dimensions and so it does not give nontrivial contribution into the equations of motion. So  the
Lagrangian density for any given spatial dimensions $D$ is sum of all Lovelock invariants (\ref{lov_lagr}) up to $n=\[\dac{D}{2}\]$ which give nontrivial contributions
into equations of motion:

\begin{equation}
{\cal L}= \sqrt{-g} \sum_n c_n L_n, \label{lagr}
\end{equation}

\noindent where $g$ is the determinant of metric tensor and
$c_n$ are coupling constants of the order of Planck length in $2n$
dimensions and summation over all $n$ in consideration is assumed. We consider the metric in the form

\begin{equation}\label{metric}
g_{\mu\nu} = \diag\{ -1, a_1^2(t), a_2^2(t),\ldots, a_n^2(t)\}.
\end{equation}

\noindent As we mentioned earlier, we are interested in the dynamics in Einstein-Gauss-Bonnet gravity, which is second order Lovelock gravity and so
 we consider $n$ up to two ($n=0$ is the boundary term while $n=1$ is Einstein-Hilbert and $n=2$ is Gauss-Bonnet contributions).
In this paper we studying the cosmological dynamics in the presence of the matter source in form of perfect fluid with an equation of state $p = \omega \rho$,
so the standard notation for its stress-energy tensor is assumed:

\begin{equation}
\begin{array}{l}
T_{\mu\nu} = (\rho + p) u_\mu u_\nu + p g_{\mu\nu}.
\end{array} \label{SET}
\end{equation}

Substituting metric (\ref{metric}) into the Lagrangian and following the usual procedure with use of (\ref{SET}) gives us the equations of motion:

\begin{equation}
\begin{array}{l}
2 \[ \sum\limits_{j\ne i} (\dot H_j + H_j^2)
+ \sum\limits_{\substack{\{ k > l\} \\ \ne i}} H_k H_l \] + 8\alpha \[ \sum\limits_{j\ne i} (\dot H_j + H_j^2) \sum\limits_{\substack{\{k>l\} \\ \ne \{i, j\}}} H_k H_l +
3 \sum\limits_{\substack{\{ k > l >  \\   m > n\} \ne i}} H_k H_l H_m H_n \] + p = 0
\end{array} \label{dyn_gen}
\end{equation}

\noindent as the $i$th dynamical equation. The first Lovelock term---the Einstein-Hilbert contribution---is in the first set of brackets, the second term---Gauss-Bonnet---is in the second set; $\alpha$
is the coupling constant for the Gauss-Bonnet contribution and we put the corresponding constant for Einstein-Hilbert contribution to unity.
Also, since we consider spatially flat cosmological models, scale
factors do not hold much in the physical sense and the equations are rewritten in terms of the Hubble parameters $H_i = \dot a_i(t)/a_i(t)$. Apart from the dynamical equations, we write down the constraint equation

\begin{equation}
\begin{array}{l}
2 \sum\limits_{i > j} H_i H_j + 24\alpha \sum\limits_{\substack{i > j >\\  k > l}} H_i H_j H_k H_l = \rho.
\end{array} \label{con_gen}
\end{equation}

As mentioned in the Introduction,
we are interested in the particular case with the scale factors split into two parts -- separately three dimensions (three-dimensional isotropic subspace), which are supposed to represent our world, and the remaining which  represent the extra dimensions ($D$-dimensional isotropic subspace). So we put $H_1 = H_2 = H_3 = H$ and $H_4 = \ldots = H_{D+3} = h$ ($D$ stands for the number of additional dimensions); then the
equations take the following form: the
dynamical equation that corresponds to $H$,

\begin{equation}
\begin{array}{l}
2 \[ 2 \dot H + 3H^2 + D\dot h + \dac{D(D+1)}{2} h^2 + 2DHh\] + 8\alpha \[ 2\dot H \(DHh + \dac{D(D-1)}{2}h^2 \) + \right. \\
 \\ \left. + D\dot h \(H^2 + 2(D-1)Hh + \dac{(D-1)(D-2)}{2}h^2 \) +
2DH^3h + \dac{D(5D-3)}{2} H^2h^2 + \right. \\
\\ \left. + D^2(D-1) Hh^3 + \dac{(D+1)D(D-1)(D-2)}{8} h^4 \] + p=0,
\end{array} \label{H_gen}
\end{equation}

\noindent the dynamical equation that corresponds to $h$,

\begin{equation}
\begin{array}{l}
2 \[ 3 \dot H + 6H^2 + (D-1)\dot h + \dac{D(D-1)}{2} h^2 + 3(D-1)Hh\] + 8\alpha \[ 3\dot H \(H^2 + \right. \right. \\
\\ \left. \left. + 2(D-1)Hh +  \dac{(D-1)(D-2)}{2}h^2 \) +  (D-1)\dot h \(3H^2 + 3(D-2)Hh + \right. \right. \\
\\  \left. \left. +
\dac{(D-2)(D-3)}{2}h^2 \) + 3H^4 +  9(D-1)H^3h + 3(D-1)(2D-3) H^2h^2 + \right. \\
\\ \left. + \dac{3(D-1)^2 (D-2)}{2} Hh^3 +   \dac{D(D-1)(D-2)(D-3)}{8} h^4 \] + p =0,
\end{array} \label{h_gen}
\end{equation}

\noindent and the constraint equation,

\begin{equation}
\begin{array}{l}
2 \[ 3H^2 + 3DHh + \dac{D(D-1)}{2} h^2 \] + 24\alpha \[ DH^3h + \dac{3D(D-1)}{2}H^2h^2 + \right. \\ \\ \left. + \dac{D(D-1)(D-2)}{2}Hh^3 +  \dac{D(D-1)(D-2)(D-3)}{24}h^4\]  = \rho.
\end{array} \label{con2_gen}
\end{equation}

Looking at (\ref{H_gen})--(\ref{con2_gen}) one can notice that the structure of the equations depends on the number of extra dimensions $D$ (terms with $(D-2)$ multiplier nullifies in $D=2$ and so on), so that we shall consider all these distinct cases separately.

Formally, the system should be complimented by the continuity equation

\begin{equation}
\begin{array}{l}
\dot\rho + (3H + Dh)(\rho + p) = 0,
\end{array} \label{cont}
\end{equation}

\noindent but the resulting system (\ref{H_gen})--(\ref{cont}) is overdetermined, so for practical purposes we skip (\ref{cont}) and use (\ref{H_gen})--(\ref{con2_gen}) for analysis.

In the previous papers, dedicated
to either vacuum or $\Lambda$-term cases in EGB or cubic Lovelock gravity~\cite{my16a, my16b, my17a, my18a, my18b, my18c}, we solved the constraint equation (\ref{con2_gen}) with respect to
either $H$ or $h$, resulting in one to three different branches of the solution with several parameters. Then the obtained expressions are substituted into (\ref{H_gen})--(\ref{h_gen}) and the system was solved with respect to $\dot H$ and $\dot h$. As a result, we have both $\dot H$ and $\dot h$ as a single-variable function with
several parameters and investigated its phase portrait to find all the asymptotic regimes, stable points etc. In the current paper, though, this method will not work, because with the
same number of independent equations we have one more variable -- the density $\rho$. So for practical purposes we use the following procedure: we substitute density $\rho$ from
(\ref{con2_gen}) into (\ref{H_gen}) and (\ref{h_gen}) with use of the equation of state $p = \omega\rho$; the resulting system is first-order with two equations and two variables and
so we can plot directional vector phase portrait to visualize the dynamics. The asymptotes as well as exact solutions are derived analytically from (\ref{H_gen})--(\ref{con2_gen}).

In the course of the paper we use the results we previously obtained in~\cite{grg10}. In particular, we demonstrated that: for $\omega < 1/3$ past GB asymptote is preserved while future asymptote is destroyed by matter and is replaced by the corresponding matter-dominated regime. For $\omega > 1/3$ the situation is opposite -- the past asymptote is destroyed and replaced by
matter-dominating regime while the future asymptote remains GB-dominated.

Before proceeding to the particular cases, let us introduce the notations we are going to use through the paper. We denote Kasner regime as $K_i$ where $i$ is the
total expansion rate in terms of the Kasner exponents $\sum p_i = (2n-1)$ where $n$ is the corresponding order of the Lovelock contribution (see, e.g.,~\cite{prd09}).
So that for Einstein-Hilbert contribution $n=1$ and $\sum p_i = 1$ (see~\cite{kasner}) and the corresponding regime is $K_1$, which is usual low-energy regime in vacuum
EGB case (see~\cite{my16a, my18a}). For Gauss-Bonnet $n=2$ and so $\sum p_i = 3$ and the regime $K_3$ is typical high-energy regime for EGB case
(again, see~\cite{my16a, my16b, my17a, my18a}). For cubic Lovelock $n=3$ and so $\sum p_i = 5$ and the regime $K_5$ is one of the high-energy regimes in that
case~\cite{my18b, my18c}.

Another power-law regime is what is called ``generalized Taub'' (see~\cite{Taub} for the original solution).
We mistakenly taken it for $K_3$ in~\cite{my16a}, but then in~\cite{my18a}
corrected ourselves and explained the details. It is a situation when for one of the subspaces the Kasner exponent
$p$ is equal to zero and for another --
to unity. So we denote
$P_{(1, 0)}$ the case with $p_H = 1, p_h = 0$ and $P_{(0, 1)}$ the case with $p_H = 0, p_h = 1$. So that in EGB for $P_{(1, 0)}$ we have $\sum p_i = 3$ (this is exactly what causes
misinterpreting of $P_{(1, 0)}$ for $K_3$) while for $P_{(0, 1)}$ we have $\sum p_i = D$.

The exponential solutions are denoted as $E$ with subindex indicating its details -- $E_{iso}$ is isotropic exponential solution and $E_{3+D}$ is anisotropic -- with different Hubble
parameters corresponding to three- and extra-dimensional subspaces. But in practice, in each particular case there are several different anisotropic exponential solutions, so that instead
of using $E_{3+D}$ we use $E_i$ where $i$ counts the number of the exponential solution ($E_1$, $E_2$ etc). In case if there are several isotropic exponential solutions, we count them with upper index:
$E_{iso}^1$, $E_{iso}^2$ etc.

And last but not least regime is what we call ``nonstandard singularity'' and denote it is as $nS$.
It is the situation which arise in Lovelock gravity due to its nonlinear nature. Since the equations
(\ref{H_gen})--(\ref{h_gen}) are nonlinear with respect to the highest derivative ($\dot H$ and $\dot h$ in our case), when we solve them, the resulting expressions are ratios with
polynomials in both numerator and denominator. So there exist a situation when the denominator is equal to zero for finite values of $H$ and\/or $h$. This situation is singular, as the
curvature invariants diverge, but it happening for finite values of $H$ and\/or $h$. Tipler~\cite{Tipler} call this kind of singularity as ``weak'' while Kitaura and
Wheeler~\cite{KW1, KW2} -- as ``type II''. Our previous research demonstrate that this kind of singularity is widely spread in EGB cosmology -- in particular, in totally anisotropic
(Bianchi-I-type) $(4+1)$-dimensional vacuum cosmological model it is the only future asymptote~\cite{prd10}.

\section{$D=1$}
\label{d1}

In this case the equations of motion (\ref{H_gen})--(\ref{con2_gen}) take form ($H$-equation, $h$-equation, and constraint correspondingly):

\begin{equation}
\begin{array}{l}
4\dot H + 6H^2 + 2\dot h + 2h^2 + 4Hh + 8\alpha \( 2(\dot H + H^2)Hh + (\dot h + h^2)H^2\) + p = 0,
\end{array} \label{D1_H}
\end{equation}

\begin{equation}
\begin{array}{l}
6\dot H + 12H^2 + 24\alpha (\dot H + H^2)H^2 + p= 0,
\end{array} \label{D1_h}
\end{equation}

\begin{equation}
\begin{array}{l}
6H^2 + 6Hh + 24\alpha H^3h = \rho.
\end{array} \label{D1_con}
\end{equation}

As we mentioned, the procedure is different from our previous papers so we are going to explain it in detail on the example of $D=1$ case. We substitute $\rho$ from (\ref{D1_con}) into
(\ref{D1_H}) and (\ref{D1_h}) with use of the equation of state $p = \omega \rho$ and solve (\ref{D1_H})--(\ref{D1_h}) with respect to $\dot H$ and $\dot h$:

\begin{equation}
\begin{array}{l}
\dot H = - H \times \dac{4\alpha\omega H^2 h + 4\alpha H^3 + \omega H + \omega h + 2H}{1 + 4\alpha H^2}, \\ \\

\dot h = - \(  48 \alpha^2 \omega H^5 h - 32\alpha^2 \omega H^4 h^2 + 16\alpha^2 H^4 h^2 + 12\alpha\omega H^4 + 8\alpha\omega H^3 h - 8\alpha \omega H^2 h^2 + 4\alpha H^4 + \right. \\ \\
\left. + 8\alpha H^2 h^2 + \omega H^2 + \omega H h - H^2 + 2Hh + h^2  \)/(1+4\alpha H^2)^2.
\end{array} \label{D1_sol1}
\end{equation}

One can see that in this case we have only one nonstandard singularity at $H = \pm 1/(2\sqrt{-\alpha}))$ and for negative $\alpha$ only. If we solve (\ref{D1_sol1}), we obtain the locations
of the exponential solutions, and they are: trivial solution $\{H=0, h=0\}$, isotropic solution $\{h\equiv H = \pm 1/\sqrt{-2\alpha}\}$, which exist only for negative $\alpha$, and
so-called ``constant-volume solution'' (CVS) (see~\cite{CST2}) with
$$
H = \pm \dac{\sqrt{-2\alpha (3\omega - 1)(\omega-1)}}{(\alpha(3\omega - 1))},~~ h=-3H.
$$
\noindent For the CVS to exist the radicand should be
positive, which happening if either $\{ \alpha > 0, \omega > 1/3 \}$ or $\{ \alpha < 0, \omega < 1/3\}$. If we substitute the CVS conditions to (\ref{D1_con}) we shall see that
only \linebreak $\{ \alpha < 0, \omega < 1/3\}$ provide positive energy density.

\begin{figure}
\centering
\includegraphics[width=1.0\textwidth, angle=0, bb= 0 0 566 566]{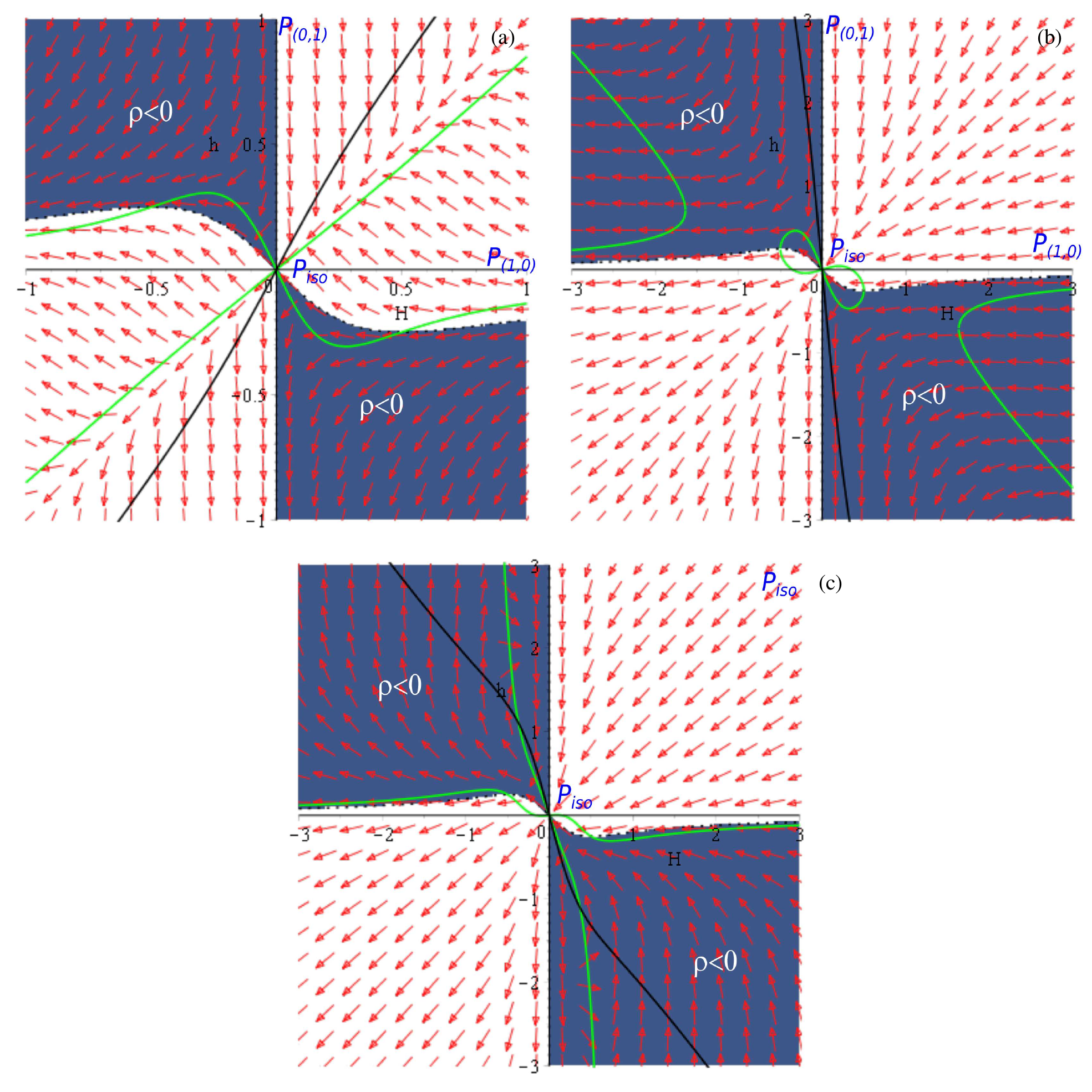}
\caption{Phase portraits for $D=1$ $\alpha > 0$ case: $\omega < 0$ on (a) panel, $0 < \omega < 1/3$ on (b) panel and $\omega > 1/3$ on (c) panel. Dark blue area corresponds to
the unphysical $\rho < 0$ initial conditions. Green curve points location of $\dot h = 0$ while black -- $\dot H = 0$
(see the text for more details).}\label{D1_1}
\end{figure}

Now with the preliminaries done, let us draw the phase portrait of the system (\ref{D1_sol1}). Despite the fact that, according to~\cite{grg10}, the fundamental change occurs at
$\omega = 1/3$,
we notice that at $\omega = 0$ there are also some minor changes which worth mentioning. So that the situation for $\alpha > 0$ is presented in Fig.~\ref{D1_1}. There we presented
$\omega < 0$ case on (a) panel, $0 < \omega < 1/3$ on (b) panel and $\omega > 1/3$ on (c) panel. Green line corresponds to $\dot h = 0$ while black to $\dot H = 0$
(see (\ref{D1_sol1})). Dark blue area corresponds to unphysical $(H, h)$ pairs with $\rho < 0$.

Let us start with $\omega < 0$ case on (a) panel and let us focus on $H > 0$ half-plane -- the situation for $H < 0$ is obviously time-reversal with
respect to $H > 0$ -- we will comment it a bit later. So that for $H > 0$ one can see that there are two past asymptotes -- $P_{(1, 0)}$ and $P_{(0, 1)}$, depending on the overall
relation between $H$ and $h$: for $H \gtrsim h$ it is $P_{(1, 0)}$ while for $H \lesssim h$ it is $P_{(0, 1)}$. As of the future asymptote, it is $P_{iso}$ -- isotropic power-law
expansion. Let us note that the future asymptote is matter-dominated while past -- GB-dominated, all well according to~\cite{grg10}. For $H < 0$, as we mentioned, the situation is
time-reversed -- so that $P_{iso}$ is past asymptote while $P_{(1, 0)}$ and $P_{(0, 1)}$ are future asymptote. Let us note that $P_{iso}$ is stable while being future asymptote but
is unstable as a past -- even a slightest deviation from isotropy turn the evolution towards either $P_{(1, 0)}$ or $P_{(0, 1)}$. So that in this case there are only two cosmological
regimes -- $P_{(1, 0)} \to P_{iso}$ and $P_{(0, 1)} \to P_{iso}$ and neither of them has realistic compactification.

The next case to consider is $0 < \omega < 1/3$ and it is presented in Fig.~\ref{D1_1}(b). One can immediately see the changes from $\omega < 0$ case (Fig.~\ref{D1_1}(a)) -- the
$\dot h = 0$ and $\dot H = 0$ curves have different structure and it slightly affects ``zeroth'' direction for $P_{(1, 0)}$ and $P_{(0, 1)}$ asymptotes -- in $\omega < 0$ case
it is $p \to 0-0$ (one can see that $p\to 0$ is reached from below) while for $0 < \omega < 1/3$ it is $p \to 0+0$ (so that it is reached from above). Except for this minor
feature there are no other differences between this and the previous cases -- the regimes are the same, so we just refer to the previous case for the description.

The last case for $\alpha > 0$ is $\omega > 1/3$ and it is presented in Fig.~\ref{D1_1}(c). One can immediately see the difference with both previous cases -- the past asymptote
switched to $P_{iso}$, and this behavior is in agreement with~\cite{grg10}. The future asymptote is the same as in the previous cases -- $P_{iso}$, so that the only cosmological
regime in this case is $P_{iso} \to P_{iso}$ -- anisotropic transition from unstable isotropic power-law regime to a stable one. Let us note that these are different isotropic
power-law solutions -- the past asymptote is matter-dominated GB while future -- matter-dominated GR so that they have different $\sum p_i$ and so different expansion rates.
This regime is unique for the matter case -- in
our previous studies dedicated to vacuum and $\Lambda$-term cases we never encountered regime of this type.

One can also notice exponential CVS in the $\rho < 0$ domain where green ($\dot h = 0$) and black ($\dot H = 0$) lines intersect. This behavior is in exact agreement with described
above and we expect CVS with positive energy density in $\{ \alpha < 0, \omega < 1/3\}$ case.

One cannot miss that the past asymptote changed at $\omega = 1/3$ while future asymptote remains the same. The reason behind it is simple -- the future asymptote in this case is
governed by GR and for GR the qualitative change similar to that is happening at $\omega = 1$ (Jacobs solution~\cite{Jacobs}). And since we limit physical equation of state within
$\omega \in (-1, 1]$, the entire allowed range is within $\omega \leqslant 1$ and so has the same regime $P_{iso}$.

\begin{figure}
\centering
\includegraphics[width=0.95\textwidth, angle=0, bb= 0 0 570 565]{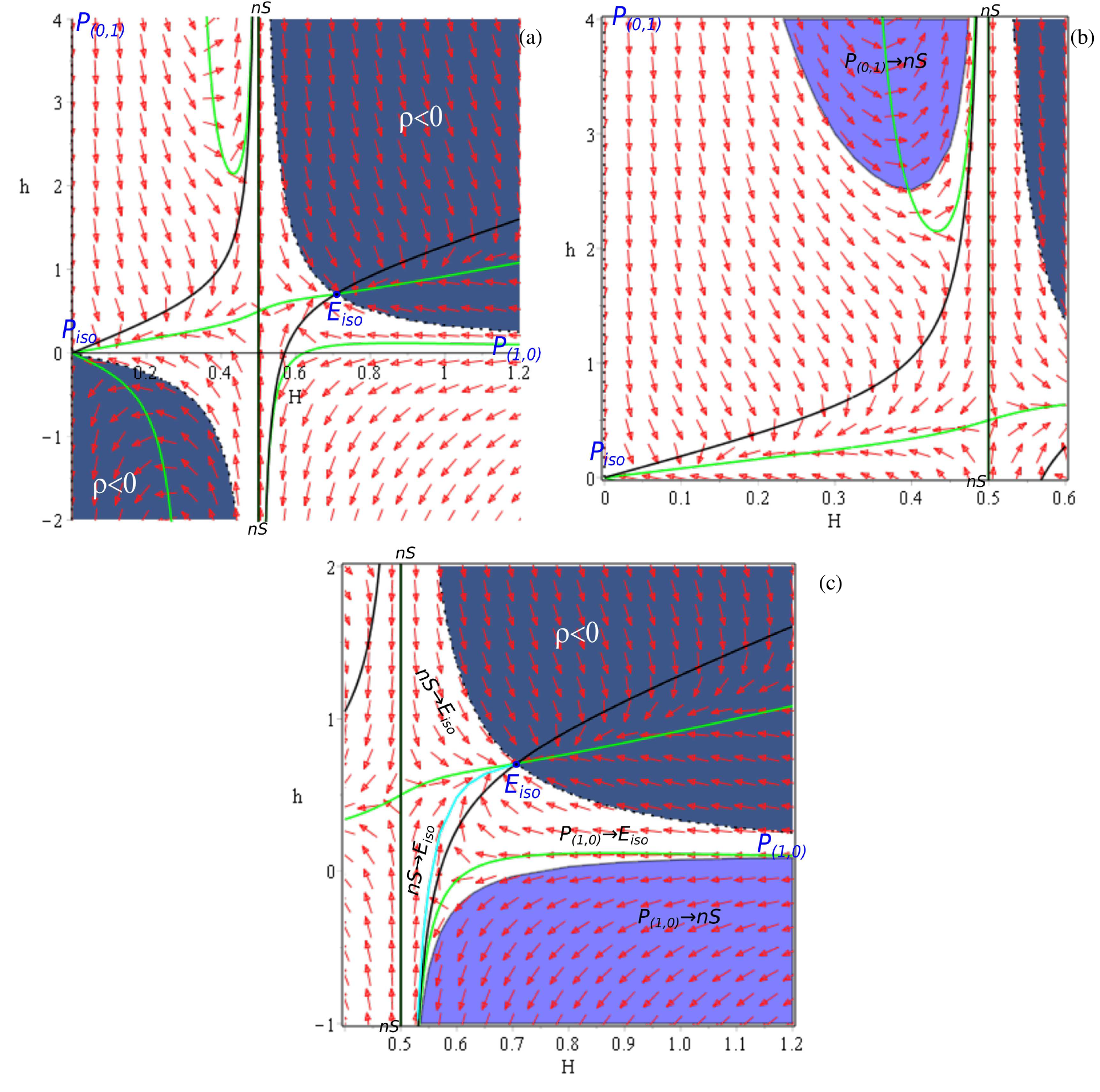}
\caption{Phase portraits for $D=1$ $\alpha < 0$, $\omega < 0$ case: general view on (a) panel, regimes for $H < H_{ns}$ on (b) panel and regimes for $H > H_{ns}$ on (c) panel.
Blue areas correspond to the initial conditions associated with specific regimes; light blue lines also separate the initial conditions associated with specific regimes.
Dark blue area corresponds to
the unphysical $\rho < 0$ initial conditions. Green curve points location of $\dot h = 0$ while black -- $\dot H = 0$
(see the text for more details).}\label{D1_2}
\end{figure}

Now let us turn to $\alpha < 0$ cases. They have richer dynamics and contain, as we demonstrated earlier, nonstandard singularities and exponential solutions. Due to this richness
we decided to present them in different figures. We start with $\alpha < 0$, $\omega < 0$ in Fig.~\ref{D1_2}. There we presented the general view of the phase portrait on (a) panel,
enlarged view of ($H > 0$, $h > 0$) up to $H < H_{nS}$ ($H_{nS}$ is the location of the nonstandard singularity ($H_{nS} = 1/(2\sqrt{-\alpha})$)) on (b) panel, and (c) panel has
focus on features at $H > H_{nS}$. As in the previous case, we limit ourselves with $H > 0$ only, as $H < 0$ regimes are time-reversal from those described for $H > 0$.
On the Fig.~\ref{D1_2}, as on the previous
Fig.~\ref{D1_1}, dark blue area corresponds to the unphysical $\rho < 0$ pairs of ($H, h$), green line depicts $\dot h = 0$ while black $\dot H = 0$. On the (a) panel we presented
the general view of the phase portrait and from it we can see that there is a fine structure for high $h$ within $H < H_{nS}$ as well as in the vicinity of $E_{iso}$ -- we presented
these areas in detail on panels (b) and (c) respectively. Apart from them, we detect $nS \to P_{iso}$ from $h < 0$, $H < H_{nS}$ area. Majority of the $h > 0$, $H < H_{nS}$
trajectories are $P_{(0, 1)} \to P_{iso}$, but at high $h$ there are $P_{(0, 1)} \to nS$ transitions, we denoted the corresponding area to blue in Fig.~\ref{D1_2}(b). So that
within $H < H_{nS}$ there are only three regimes: $nS \to P_{iso}$, $P_{(0, 1)} \to nS$ and $P_{(0, 1)} \to P_{iso}$.

The $H > H_{nS}$ area is shown in detail in Fig.~\ref{D1_2}(c). We can see that at high $h$ the regime is $nS \to E_{iso}$, while the situation at low and negative $h$ is more
complicated. At high $H$ the past asymptote is $P_{(1, 0)}$ and it gives rise to two regimes -- $P_{(1, 0)} \to E_{iso}$ and $P_{(1, 0)} \to nS$, the latter is colored in blue on
panel (c). The final regime is another $nS \to E_{iso}$, which originate from the same $nS$ but at $h\to -\infty$. The boundary between $P_{(1, 0)} \to E_{iso}$ and $nS \to E_{iso}$
is denoted by light blue line in Fig.~\ref{D1_2}(c).

So that for $\alpha < 0$, $\omega < 0$ there are seven different regimes, and among them six are distinct: $nS \to P_{iso}$, $P_{(0, 1)} \to nS$ and $P_{(0, 1)} \to P_{iso}$ at
$H < H_{nS}$ and $P_{(1, 0)} \to E_{iso}$, $P_{(1, 0)} \to nS$ and $nS \to E_{iso}$. The last mentioned regime exists in two variations -- its $nS$ could have either $h\to +\infty$ or
$h\to -\infty$.

Before moving on to the next case, it is important to mention the CVS solution -- we theoretically predicted its existence but it has not participated in the regimes description.
In reality it
exists at $h < 0$, $H > H_{nS}$ where green ($\dot h = 0$) and black ($\dot H = 0$) lines intersect. For $\omega < 0$ the intersection angle is very small so on a chosen scales
it is very hard to locate it -- in the next case the situation improves and we are going to describe CVS in detail.

\begin{figure}
\centering
\includegraphics[width=0.95\textwidth, angle=0, bb= 0 0 571 567]{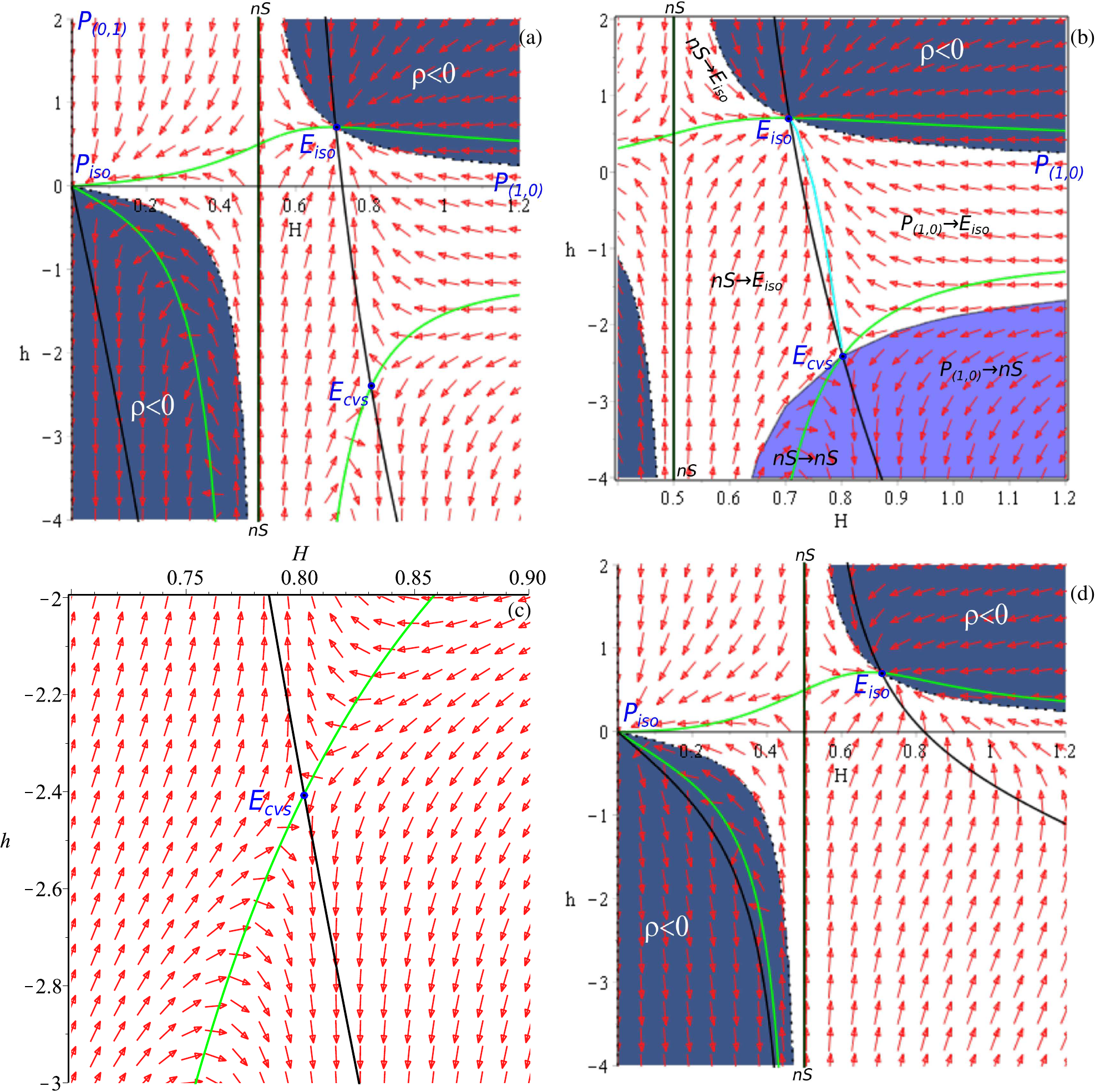}
\caption{Phase portraits for $D=1$ $\alpha < 0$, $\omega > 0$ case: $0 < \omega < 1/3$ general view on (a) panel, $0 < \omega < 1/3$ regimes for $H > H_{ns}$ on (b) panel,
vicinity of CVS on (c) panel and $\omega > 1/3$ general view on (d) panel.
Blue area corresponds to the initial conditions associated with specific regime. Light blue line separate initial conditions leading to different regimes.
Dark blue area corresponds to
the unphysical $\rho < 0$ initial conditions. Green curve points location of $\dot h = 0$ while black -- $\dot H = 0$
(see the text for more details).}\label{D1_3}
\end{figure}

Now let us move on to $\alpha < 0$, $1/3 > \omega > 0$ which we presented in Fig.~\ref{D1_3}(a)--(c). The notations are the same as in the previous figures: dark blue area
corresponds to the unphysical $\rho < 0$ pairs of ($H, h$), green line depicts $\dot h = 0$ while black $\dot H = 0$. On (a) panel we presented general view of the phase portrait,
on (b) panel we focus on the features in the vicinity of the isotropic exponential solution $E_{iso}$ and on (c) panel we illustrate constant-volume exponential solution (CVS).
From Fig.~\ref{D1_3}(a) we can see that for $H < H_{nS}$ the regimes are $P_{(0, 1)} \to P_{iso}$ for $h > 0$ and $nS \to P_{iso}$ for $h < 0$. The difference in $P_{(0, 1)}$ between
$\omega < 0$ and $1/3 > \omega > 0$ is the same as described for $\alpha > 0$ case -- for $\omega < 0$ we have $p_H \to 0-0$ while for $1/3 > \omega > 0$ it is $p_H \to 0+0$;
the same is true about $p_h$ for $P_{(1, 0)}$. The dynamics for $H > H_{nS}$ is more abundant: the past asymptotes are $nS$ at $H = H_{nS}$ (note that there are two of them -- at
$h\to +\infty$ and $h\to -\infty$) and $P_{(1, 0)}$ at $H\to\infty$. So that the regimes are (see Fig.~\ref{D1_3}(b)): $nS \to E_{iso}$ (formally two different, as there are two
$nS$ at $h\to\pm\infty$), $P_{(1, 0)}\to E_{iso}$ and $P_{(1, 0)} \to nS$ (the latter is represented as blue area) and $nS\to nS$ as a small part of blue area, roughly separated by
black line. One can see that of all regimes only $P_{(0, 1)} \to P_{iso}$ and $P_{(1, 0)} \to E_{iso}$ are nonsingular, but neither of them have realistic compactification.

In this case we can clearly see CVS, it is represented as $E_{CVS}$. One can also see that despite existent, it is neither past nor future asymptote for any regime. From
Fig.~\ref{D1_3}(c) one can clearly see why is it -- it is a pole on the phase portrait, referring to the marginal stability of the solution. It is expectant -- while studying
the stability of the exponential solutions in~\cite{my15} (and later it was generalized in~\cite{iv16}), the stability analysis works for $\sum H_i \ne 0$. On the contrary,
CVS has, as its name assumes, $\sum H_i = 0$, making the standard stability analysis unviable, which could be treated as a marginal stability, and now we exactly demonstrate it.
For vacuum and $\Lambda$-term cases in EGB as well as in the cubic Lovelock gravity we also have CVS, but due to the lesser number of the degrees of freedom, CVS in these cases have
directional stability (i.e. stable from $\sum H_i > 0$ direction and unstable from $\sum H_i < 0$ direction, see~\cite{my16b, my17a, my18a, my18b, my18c}).

Finally let us consider $\alpha < 0$, $\omega > 1/3$ case, which is presented in Fig.~\ref{D1_3}(d). The dynamics in this case is much more simple then in previous cases -- for both
$H < H_{nS}$ the only regime is $nS \to P_{iso}$ for both $h > 0$ and $h < 0$ while for $H > H_{nS}$ the only regime is $nS \to E_{iso}$ for both $h > 0$ and $h < 0$
So that there are neither nonsingular nor realistic regimes in this case.

To conclude $D=1$ case with matter in form of perfect fluid, we have three nonsingular regimes for $\alpha > 0$: $P_{(1, 0)} \to P_{iso}$, $P_{(0, 1)} \to P_{iso}$ and
$P_{iso} \to P_{iso}$, but neither of them have realistic compactification. Let us note that there are no other regimes apart from the just mentioned, so that all possible regimes
in $\alpha >  0$ case are nonsingular. The situation with $\alpha < 0$ is more complicated and we have $P_{(0, 1)} \to P_{iso}$ and $P_{(1, 0)} \to E_{iso}$ as nonsingular regimes; they
exist for $\omega < 1/3$ and for $\omega > 1/3$ there are no nonsingular regimes. Nevertheless, neither of nonsingular regimes for $\alpha < 0$ have realistic compactification
either, making $D=1$ case degenerative with respect to realistic compactifications.

\section{$D=2$}
\label{d2}

In this case the equations of motion (\ref{H_gen})--(\ref{con2_gen}) take form ($H$-equation, $h$-equation, and constraint correspondingly):

\begin{equation}
\begin{array}{l}
4\dot H + 6H^2 + 4\dot h + 6h^2 + 8Hh + 8\alpha \( 2(\dot H + H^2)(2Hh + h^2) + \right. \\ \\ + \left. 2(\dot h + h^2)(H^2+2Hh) + 3H^2 h^2   \) + p = 0,
\end{array} \label{D2_H}
\end{equation}

\begin{equation}
\begin{array}{l}
6\dot H + 12H^2 + 2\dot h + 2h^2 + 6Hh + 8\alpha \( 3(\dot H + H^2)(H^2 + 2Hh ) + \right. \\ \\ + \left. 3(\dot h + h^2)H^2 + 3H^3 h   \) + p = 0,
\end{array} \label{D2_h}
\end{equation}

\begin{equation}
\begin{array}{l}
6H^2 + 12Hh + 2h^2 + 24\alpha \( 2H^3h + 3H^2 h^2\) = \rho.
\end{array} \label{D2_con}
\end{equation}

Similar to the previously described procedure, we substitute $\rho$ from (\ref{D2_con}) into
(\ref{D2_H}) and (\ref{D2_h}), use the equation of state $p = \omega \rho$ and solve (\ref{D2_H})--(\ref{D2_h}) with respect to $\dot H$ and $\dot h$:

\begin{equation}
\begin{array}{l}
\dot H = -  \dac{P_1}{4Q}, \quad \dot h = -  \dac{P_2}{4Q}, ~~\mbox{where} \\
Q = 24\alpha^2 H^4 + 48\alpha^2 H^3 h + 72\alpha^2 H^2 h^2 + 6\alpha H^2 + 20\alpha Hh - 2\alpha h^2 + 1, \\
P_1 = - 96 \alpha^2 \omega H^5 h + 240 \alpha^2 \omega H^4 h^2 + 576 \alpha^2 \omega H^3 h^3 + 96 \alpha^2 H^6 + 288 \alpha^2 H^5 h + 336 \alpha^2 H^4 h^2 - \\
- 12\alpha \omega H^4 +
48 \alpha \omega H^3 h + 128 \alpha \omega H^2 h^2 + 16\alpha\omega Hh^3 + 36\alpha H^4 + 128\alpha H^3 h + 16\alpha H^2 h^2 + \\ + 3\omega H^2 + 6\omega Hh + \omega h^2 + 9H^2
+2Hh- h^2, \\
P_2 = 288 \alpha^2 \omega H^5 h + 624 \alpha^2 \omega H^4 h^2 + 96\alpha^2 \omega H^3 h^3 - 288 \alpha^2 \omega H^2 h^4 + 48\alpha^2 H^4 h^2 + \\ + 384 \alpha^2 H^3 h^3 +
288\alpha^2 H^2 h^4 + 36\alpha\omega H^4 + 120 \alpha\omega H^3 h + 72\alpha\omega H^2 h^2 - 40\alpha\omega Hh^3 - 8\alpha\omega h^4 + \\ + 12\alpha H^4 +  96\alpha H^2 h^2
+ 80\alpha Hh^3 - 8\alpha h^2 + 3\omega H^2 + 6\omega Hh + \omega h^2 - 3H^2 + 6Hh + 7h^2.
\end{array} \label{D2_sol1}
\end{equation}

The positions of the nonstandard singularities are determined by zeros of $Q$ and one can see that now, unlike $D=1$ case, they are not just straight lines. One can also see that
$Q$ is quadratic with respect to $h$ so one can easily solve and plot it. Calculating the discriminant of $Q$ with respect to $h$ gives us

$$
\mathcal{D}_h = -8\alpha \( 4\alpha H^2 - 1\) \( 1+12\alpha H^2 \)^2.
$$

\noindent From it one can easily see that for $\alpha > 0$ discriminant is positive if $H^2 < 1/(4\alpha)$ while for negative $\alpha$ the discriminant is always negative -- so
that for negative $\alpha$ there are no nonstandard singularities -- very unexpected result, considering that for $D=1$ nonstandard singularities exist only for negative~$\alpha$.

The exponential solutions
are defined by $P_1 = P_2 \equiv 0$ and are as follows: trivial solution $\{H=0, h=0\}$, isotropic solution $\{h\equiv H = \pm 1/\sqrt{-6\alpha}\}$, which exist only for
negative $\alpha$, CVS which we describe below and one more solution with

\begin{equation}
\begin{array}{l}
192 \eta^3 - 112 \eta^2 + 4\eta - 1 = 0~~\mbox{where}~~ \eta = \alpha H^2, \\
h = - H \dac{P_3}{P_4},~~\mbox{where} \\
P_3 = 225340272 \eta^2 \omega^3 - 250213488 \eta^2 \omega^2 - 4635576 \eta \omega^3 + 103144144\eta^2 \omega + 5149104 \eta \omega^2 + \\ + 2085579 \omega^3 - 12914000\eta^2
- 2123896 \eta\omega - 2315661 \omega^2 + 265760\eta + 954661\omega - 119555, \\
P_4 = 312054624 \eta^2\omega^3 - 346496736\eta^2\omega^2 - 6419688\eta\omega^3 + 142831136\eta^2\omega + 7129752\eta\omega^2 + \\ + 2888040\omega^3 - 17883040\eta^2 -
2938904\eta\omega - 3206880\omega^2 + 367720\eta + 1322024\omega - 165520.
\end{array} \label{D2_sol2}
\end{equation}

Solving equation for $\eta$ we obtain single root $\eta = \sqrt[3]{10}/9 + \sqrt[3]{100}/36 + 7/36 \approx 0.56276$ and the corresponding solution for $H$ exists only for
$\alpha > 0$. Then substituting the obtained root for $\eta$ into $h$, we find out that the resulting ratio is almost constant within $\omega \in [-1, 1]$ and
$h/H \approx -0.72212$. So that this anisotropic exponential solution have different signs for $H$ and $h$ (and so it could describe realistic compactification) and it
exist for $\alpha > 0$ for all $\omega$ except for removable singularity at $\omega = 13/27 - 2\sqrt[3]{100}/27 + \sqrt[3]{10}/27 \approx 0.21745$, found from zeroth of $P_4$.

The mentioned constant-volume solution has

$$
H^2 = (\omega-1)/(4\alpha(3\omega - 1))
$$

\noindent and from the requirement of the positivity it exists in two domains:
$(\alpha > 0, \omega < 1/3)$ and \linebreak $(\alpha < 0, \omega > 1/3)$. Substituting the solution into the integral of energy gives us

$$
\rho = - \dac{15(\omega - 1)}{4\alpha (3\omega - 1)^2},
$$

\noindent so that only $(\alpha > 0, \omega < 1/3)$ domain gives CVS with positive energy density.

\begin{figure}
\centering
\includegraphics[width=0.95\textwidth, angle=0, bb= 0 0 571 567]{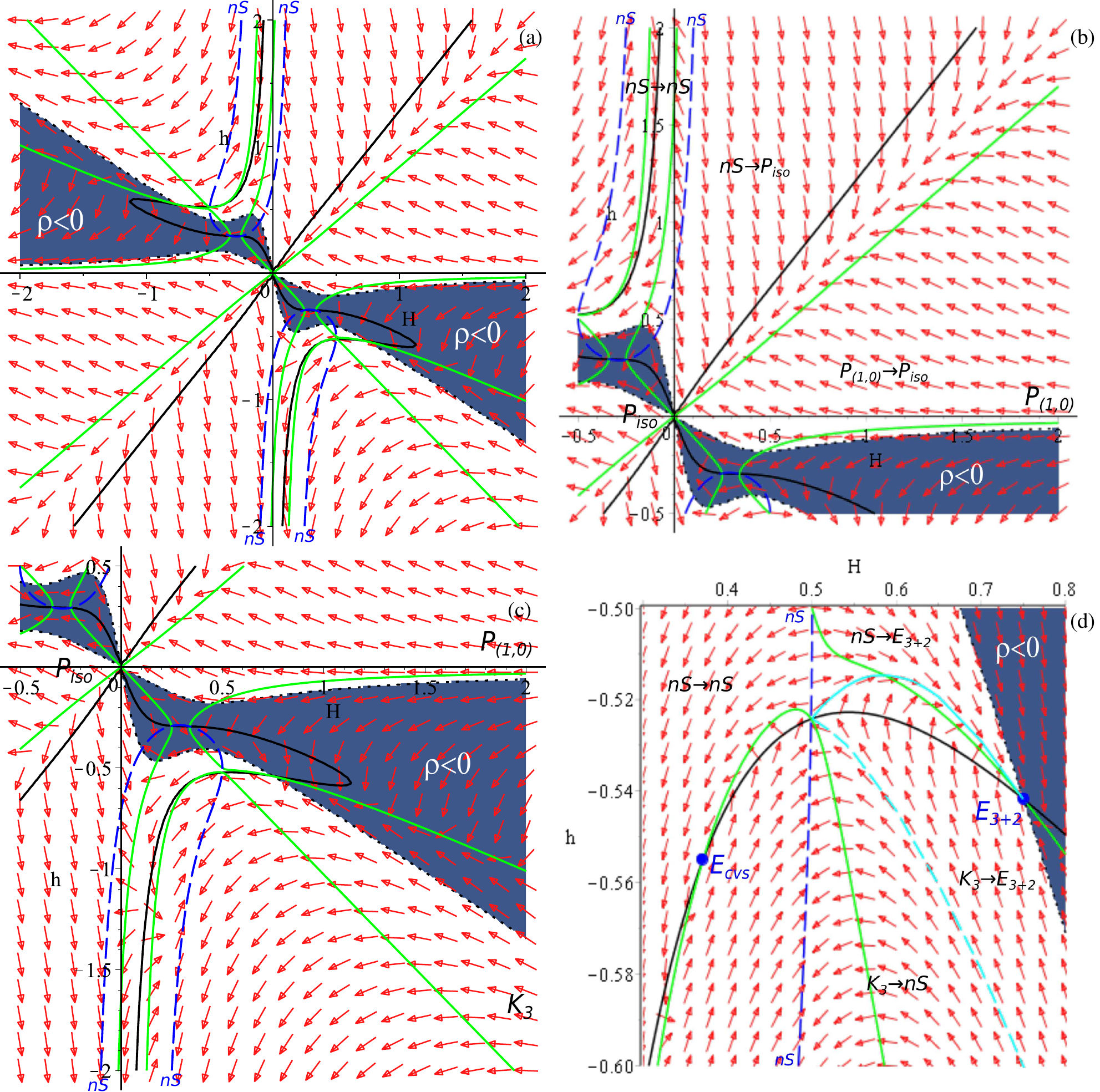}
\caption{Phase portraits for $D=2$ $\alpha > 0$, $\omega < 0$ case: general view on (a) panel, focus on $h > 0$ regimes on (b) panel, focus on $h < 0$ regimes on (c) panel and
focus on the regimes in the vicinity of $E_{3+2}$ solution on (d) panel.
 Light blue line separate initial conditions leading to different regimes.
Dark blue area corresponds to
the unphysical $\rho < 0$ initial conditions. Green curve points location of $\dot h = 0$, black -- $\dot H = 0$ and dashed blue -- to nonstandard singularities
(see the text for more details).}\label{D2_1}
\end{figure}

\begin{figure}
\centering
\includegraphics[width=0.95\textwidth, angle=0, bb= 0 0 571 567]{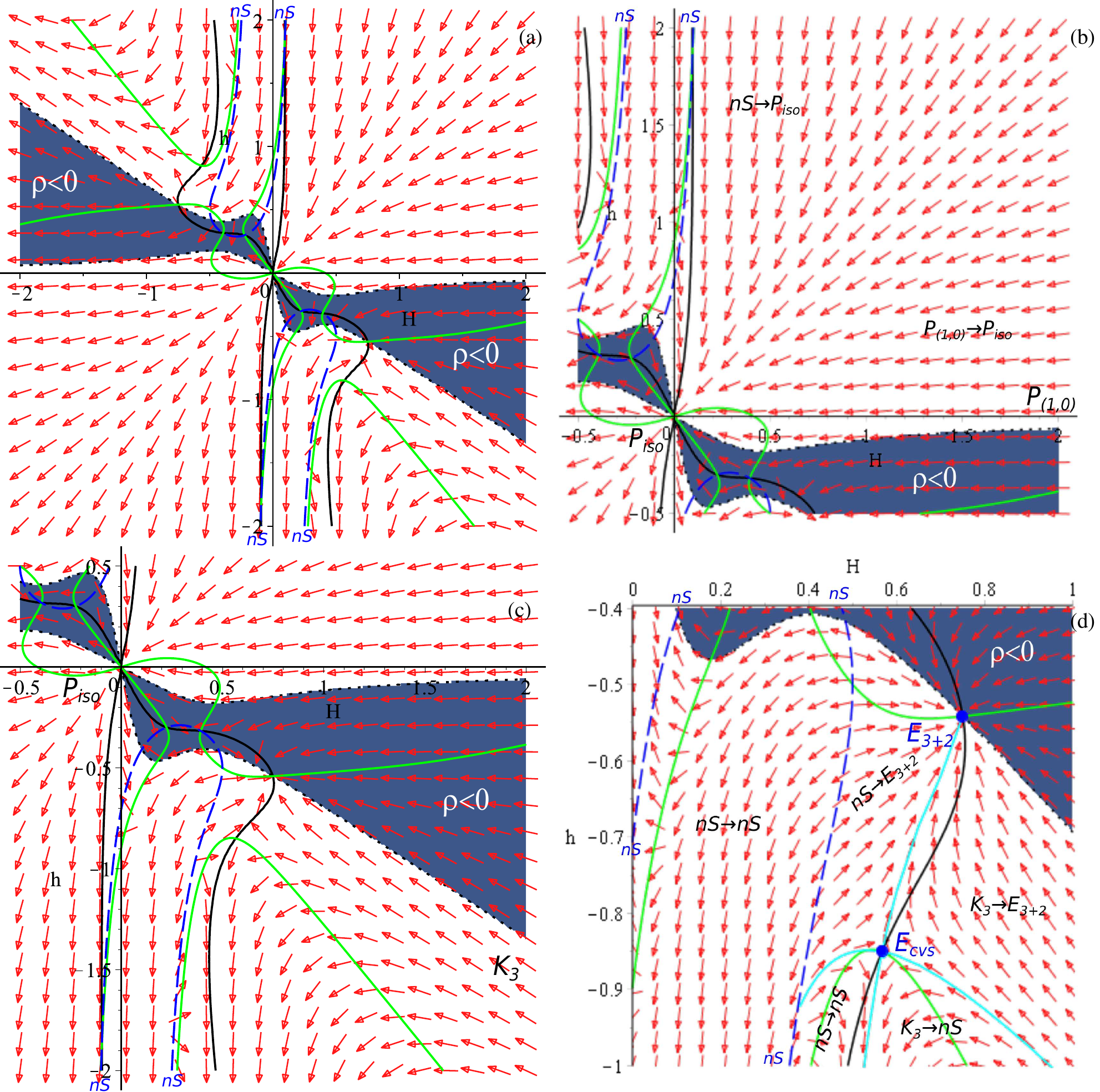}
\caption{Phase portraits for $D=2$ $\alpha > 0$, $1/3 > \omega > 0$ case: general view on (a) panel, focus on $h > 0$ regimes on (b) panel, focus on $h < 0$ regimes on (c) panel and
focus on the regimes in the vicinity of $E_{3+2}$ solution on (d) panel.
 Light blue line separate initial conditions leading to different regimes.
Dark blue area corresponds to
the unphysical $\rho < 0$ initial conditions. Green curve points location of $\dot h = 0$, black -- $\dot H = 0$ and dashed blue -- to nonstandard singularities
(see the text for more details).}\label{D2_2}
\end{figure}

\begin{figure}
\centering
\includegraphics[width=0.95\textwidth, angle=0, bb= 0 0 568 565]{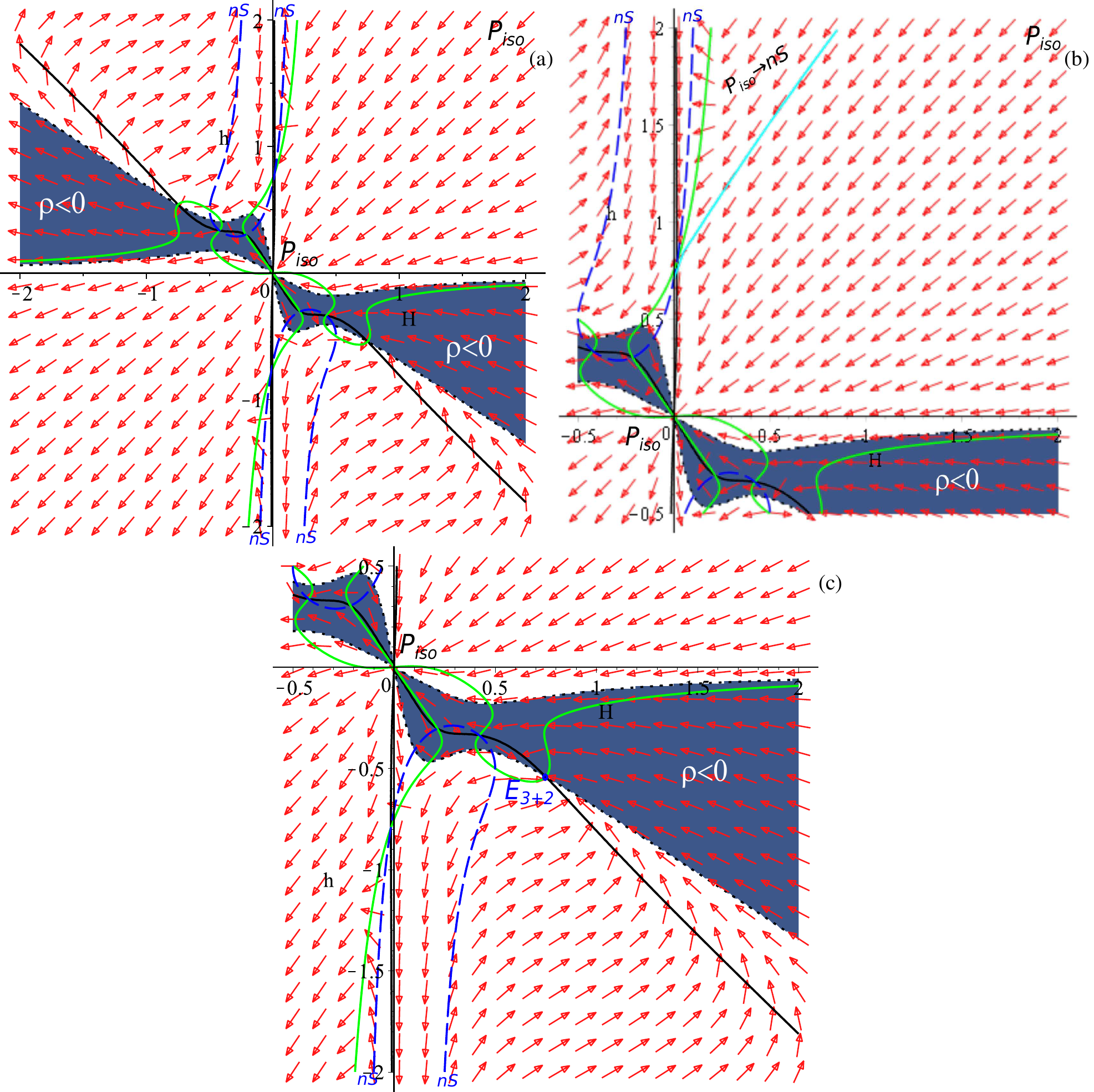}
\caption{Phase portraits for $D=2$ $\alpha > 0$, $\omega > 1/3$ case: general view on (a) panel, focus on $h > 0$ regimes on (b) panel and focus on $h < 0$ regimes on (c) panel.
 Light blue line separate initial conditions leading to different regimes.
Dark blue area corresponds to
the unphysical $\rho < 0$ initial conditions. Green curve points location of $\dot h = 0$, black -- $\dot H = 0$ and dashed blue -- to nonstandard singularities
(see the text for more details).}\label{D2_3}
\end{figure}

With the preliminaries done, let us present the resulting phase portraits. From the structure of the nonstandard singularities and exponential solutions one can see that $D=2$ case
is more complicated than the previous $D=1$ and so the structure of the regimes. We presented $\alpha > 0$ cases in Figs.~\ref{D2_1}--\ref{D2_3}.

The first case to consider is $\alpha > 0$, $\omega < 0$, presented in Fig.~\ref{D2_1}. On (a) panel we presented general view of the phase portrait. One can see the difference
from $D=1$ cases -- now we have two branches and they are separated by $\rho < 0$ unphysical region, colored in dark blue. As the dynamics is more complicated, it is natural to
focus on different regions of the map and mark different features on them. This way we focus on $h > 0$ half-plane on (b) panel. One can immediately see the difference with
$D=1$ case -- in the latter we have $P_{(0, 1)}$ as a past asymptote at $h\to +\infty$, $H\to 0$ while now in $D=2$ it is replaced by nonstandard singularity. So that the regimes
are $nS \to P_{iso}$ for $h\gtrsim H$ and $P_{(1, 0)} \to P_{iso}$ for $h\lesssim H$; the latter is the same as in $D=1$ case.

On (c) panel we presented the phase portrait focused on $h<0$ half-plane. Despite the fact that the very ``upper'' part (above the $\rho < 0$ separation region) formally has $h<0$,
dynamically it belongs to $h>0$, so we omit it from the description here. The $h<0$ half-plane also has ``double'' nonstandard singularity, as well as $K_3$ as high-energy past
asymptote. The dynamics in the vicinity of the exponential solutions is presented on (d) panel in details. One can see that we have two exponential solutions here -- constant-volume
solution (CVS) as well as anisotropic exponential solution $E_{3+2}$ with three expanding ($H > 0$) and two contracting ($h<0$) dimensions. The CVS has the same property as in
$D=1$ -- it is a pole and so is not participating in the dynamical evolution. So that the regimes are (see Fig.~\ref{D2_1}(d)): $K_3 \to E_{3+2}$ and $K_3 \to nS$, separated
by dashed
blue line, both originate from $K_3$ -- in $D=1$ there is no such high-energy regime. The remaining two regimes originate from $nS$: $nS \to nS$ (located between two $nS$) and
$nS \to E_{3+2}$. One can see that of these regimes $K_3 \to E_{3+2}$ is both nonsingular and has realistic compactification. But its area of emergence is quite small with respect to
the total initial conditions area.

So that in $\alpha > 0$, $\omega < 0$ case we have a number of regimes with two of them -- $P_{(1, 0)} \to P_{iso}$ and $K_3 \to E_{3+2}$ being nonsingular and the latter in addition
has realistic compactification.

The next case is $\alpha > 0$, $1/3 > \omega > 0$, presented in Fig.~\ref{D2_2}. The general panels layout is the same as in Fig.~\ref{D2_1} -- general view on (a) panel, focus on
$h > 0$ on (b) panel, focus on $h < 0$ on (c) panel and focus on the physical regimes on (d) panel. On the $h > 0$ half-plane the difference from the previous $\omega < 0$ case is
quite similar to that in $D=1$ case -- the asymptote regimes are the same but they are approached with the different limits: for $P_{(1, 0)}$ we have $p_H \to 1-0$ at $\omega < 0$
and $p_H \to 1+0$ at $1/3 > \omega > 0$. So that the regimes at $h > 0$ are the same as in $\omega < 0$ case: $nS \to P_{iso}$ for $h\gtrsim H$ and $P_{(1, 0)} \to P_{iso}$ for
$h\lesssim H$.

For $h < 0$ the situation is also similar to $\omega < 0$ case and the regimes are also the same. The difference lies in the abundance of the regimes -- comparing
Fig.~\ref{D2_1}(d) with Fig.~\ref{D2_2}(d) one can see that the measure of the $K_3 \to E_{3+2}$ regime has grown. Additional investigation reveals that the maximal measure is
achieved at $\omega \to 1/3-0$ when it fill the entire area bounded by the green line, then black line and $\rho = 0$. The other regimes -- $K_3 \to nS$, $nS \to nS$ and
$nS \to K_{3+2}$ also changing their bounds of existence, but since they are non-physical, they are of little interest to us.

To conclude, in $\alpha > 0$, $1/3 > \omega > 0$ case we have the same regimes as in $\omega < 0$, but the measure if the physically realistic $K_3 \to E_{3+2}$ regime is growing
with increase of $\omega$ until it reach its maximum at $\omega \to 1/3-0$.

The final case for $\alpha > 0$ is $\omega > 1/3$, presented in Fig.~\ref{D2_3}. Similar to the previous cases, the panels layout is as follows: on (a) panel we presented general
view of the phase portrait, focus on $h > 0$ is done on (b) panel and focus on $h < 0$ is on (c) panel; unlike previous cases there is no ``fine-structure'' of the regime in the
vicinity of $E_{3+2}$. Similar to $D=1$, at $\omega=1/3$ we have a change of past asymptote -- now it is $P_{iso}$, so that on $h > 0$ half-plane the main regime is
$P_{iso} \to P_{iso}$ -- anisotropic transition between two different isotropic power-law regimes -- the same regime we have in $D=1$, $\omega > 1/3$ case.
In the $h < 0$ half-plane we also have change of past asymptote -- now it is solely nonstandard singularity, so that the only regimes for $h < 0$ are $nS \to nS$ and
$nS \to E_{3+2}$ (see Fig.~\ref{D2_3}).

To conclude $D=2$ $\alpha > 0$ regimes, we have two nonsingular regimes $P_{(1, 0)} \to P_{iso}$ and $K_3 \to E_{3+2}$ at $\omega < 1/3$ with the latter having realistic
compactification; its measure over the total initial conditions space is growing with the increase of $\omega$ until reaching its maxima at $\omega \to 1/3-0$. For $\omega > 1/3$
the only non-singular regime is $P_{iso} \to P_{iso}$.

Now let us turn our attention to the $\alpha < 0$ regimes, which are a bit more difficult to attend. Let us have a look on (\ref{D2_con}) and calculate its discriminant with
respect to $h$:

\begin{equation}
\begin{array}{l}
\mathcal{D}_h = 2304\alpha^2 x^3 - 576\alpha x^2 + 288\alpha\rho x + 96 x + 8\rho,
\end{array} \label{D2_discr}
\end{equation}

\noindent where $x=H^2$. One can check that for $\alpha > 0$ (and $\rho \geqslant 0$ of course) $\mathcal{D}_h > 0$ always, so that $\forall \rho > 0 \,\exists \,h_{1, 2}$ which are solutions of
(\ref{D2_con}).
 On the contrary, for $\alpha < 0$ the situation is different: $\forall \,H \,\exists\, \rho_0$ such as for $\rho > \rho_0$ we have
$\mathcal{D}_h < 0$. This means that for $\alpha < 0$ there exists maximal value for density

\begin{figure}
\centering
\includegraphics[width=0.95\textwidth, angle=0, bb= 0 0 567 564]{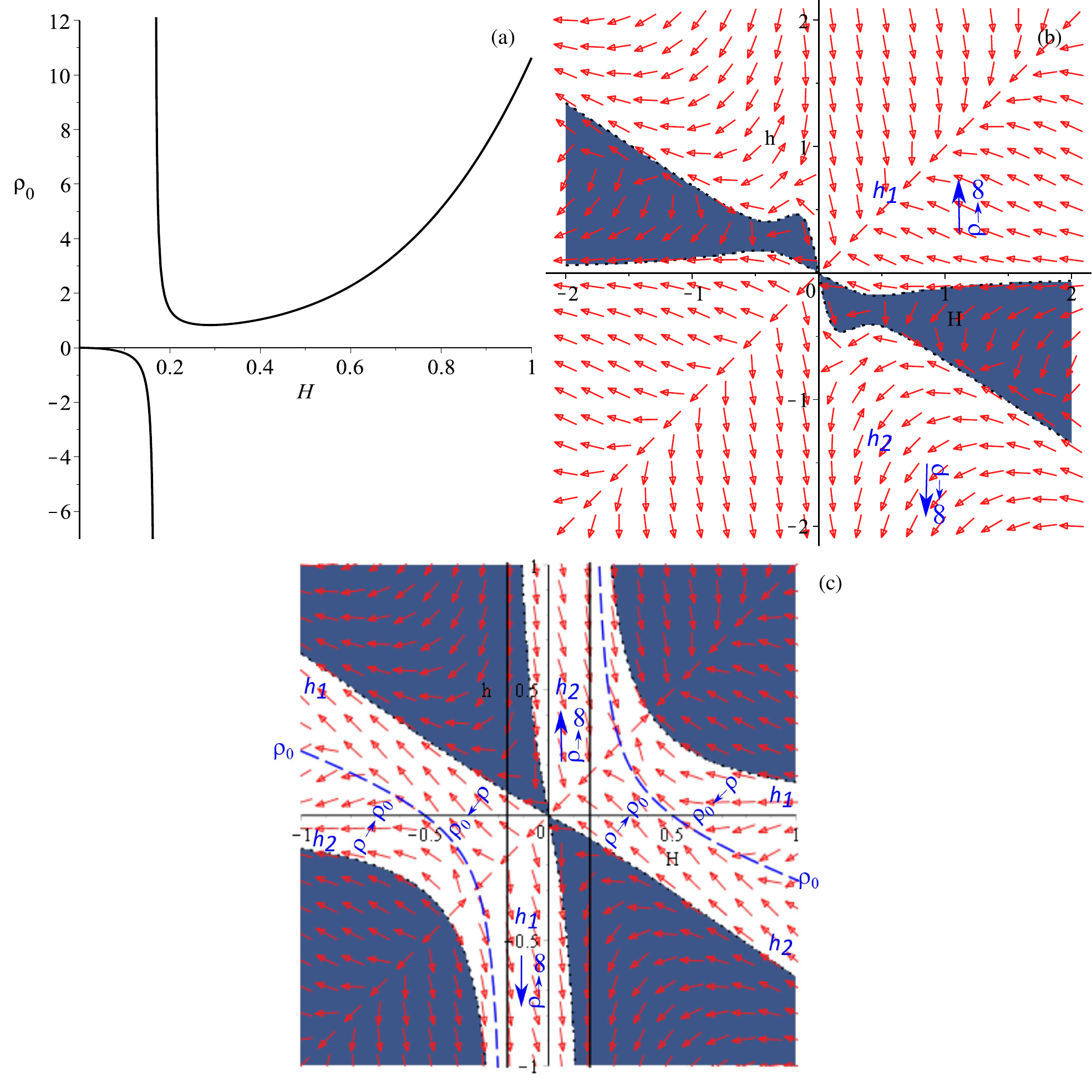}
\caption{Maximal density $\rho_0$ and its influence on the branches in $D=2$: $\rho_0$ as a function of $H$ on (a) panel, infinite maximal density in the $\alpha > 0$ case on
(b) panel, $\alpha < 0$ case on (c) panel: no maximal density for $H^2 < -1/(36\alpha)$ and separation of the phase space by $\rho=\rho_0(H, h)$ curve for $H^2 > -1/(36\alpha)$
(see the text for more details).}\label{D2_4}
\end{figure}

\begin{equation}
\begin{array}{l}
\rho_0 = - 12H^2 \dac{24\alpha^2 H^4 - 6\alpha H^2 + 1}{36\alpha H^2 + 1};
\end{array} \label{D2_rho0}
\end{equation}

\noindent we presented the plot of this relation in Fig.~\ref{D2_4}(a). From the graph and from the expression one can see that for $H^2 < -1/(36\alpha)$ the maximal density
formally is negative, but (\ref{D2_discr}) nevertheless is positive, so that the limit is absent. For comparison reasons we presented $\alpha > 0$ case in Fig.~\ref{D2_4}(b) --
one can easily see that $\rho$ is unlimited there, as there are neither upper limit for $h_1$ nor lower limit for $h_2$. The same situation is for $\alpha < 0$ and
$H^2 < -1/(36\alpha)$, presented in Fig.~\ref{D2_4}(c) between vertical black lines. The very lines correspond to $H^2 = -1/(36\alpha)$ which are nonstandard singularities for
the vacuum ($\rho = 0$) case -- and one can see that in the limits $h\to\pm\infty$ with the appropriate sign, the vacuum regime is reached.
For $H^2 > -1/(36\alpha)$ the situation changes, as $\rho_0$ from (\ref{D2_rho0}) starts to be positive. The line for $\rho=\rho_0$ is represented by
blue dashed line and the density increases from $\rho=0$ (the boundary of the dark blue region) towards blue dashed line. We shall discuss this feature -- the existence of the
maximal density -- a bit later and now let us return to the $\alpha < 0$ regimes.

\begin{figure}
\centering
\includegraphics[width=0.95\textwidth, angle=0, bb= 0 0 566 566]{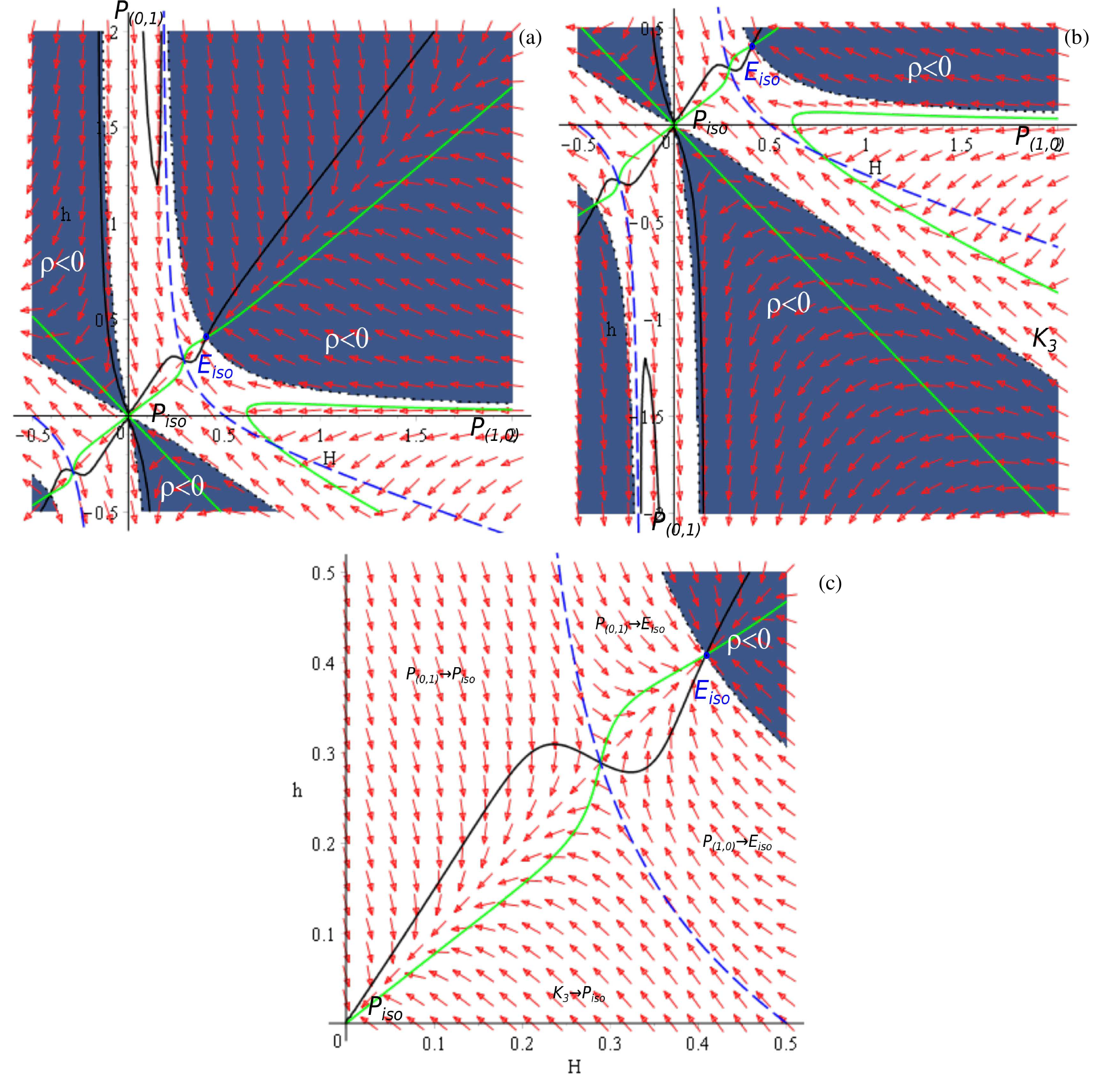}
\caption{Phase portraits for $D=2$ $\alpha < 0$, $\omega < 0$ case: focus on $h > 0$ regimes on (a) panel, focus on $h < 0$ regimes on (b) panel and the central area between
$P_{iso}$ and $E_{iso}$ on (c) panel.
 Dashed blue line corresponds to the $\rho_0$ from (\ref{D2_rho0}).
Dark blue area corresponds to
the unphysical $\rho < 0$ initial conditions. Green curve points location of $\dot h = 0$ while black -- $\dot H = 0$
(see the text for more details).}\label{D2_5}
\end{figure}

\begin{figure}
\centering
\includegraphics[width=0.95\textwidth, angle=0, bb= 0 0 568 566]{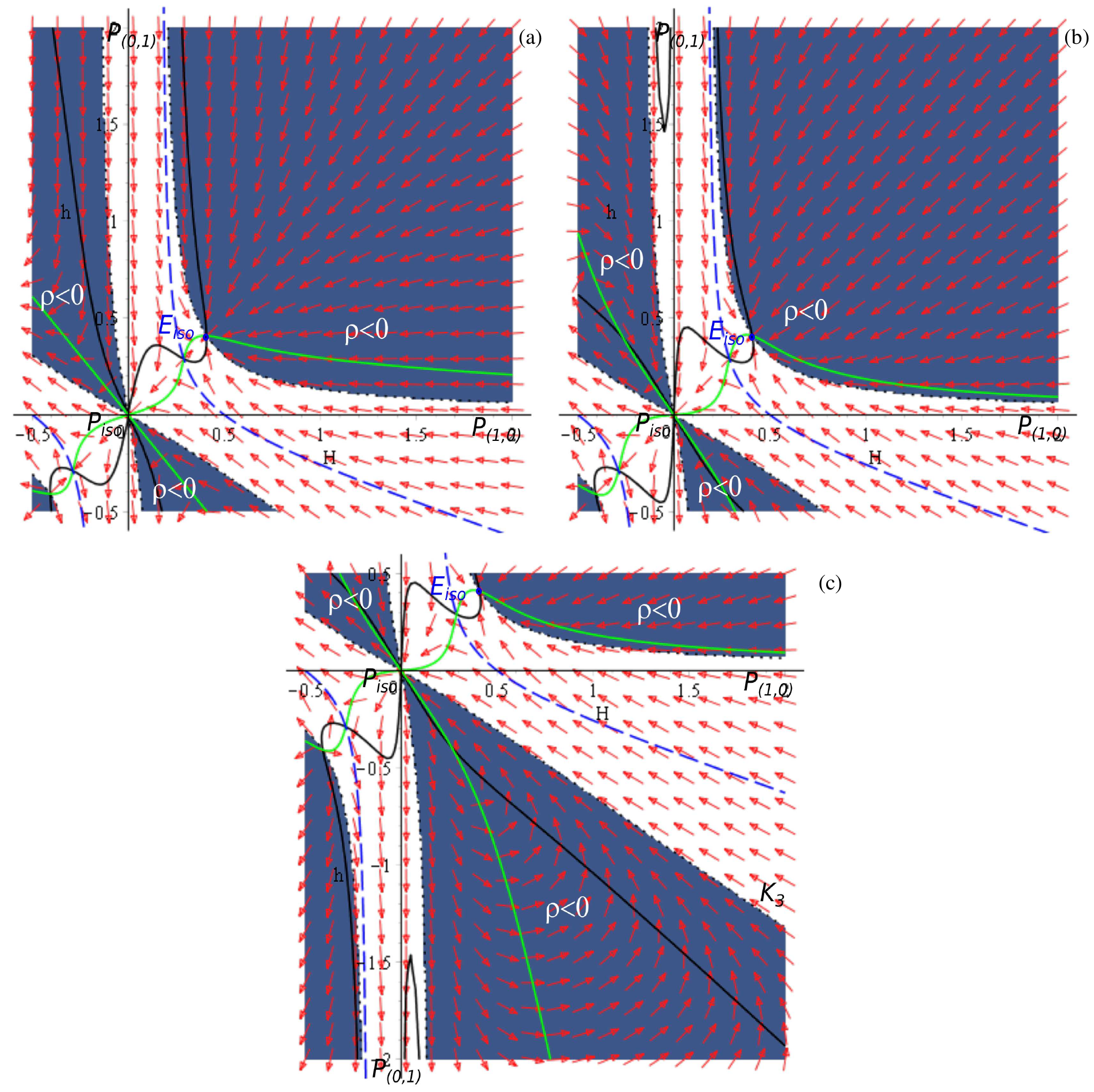}
\caption{Phase portraits for $D=2$ $\alpha < 0$, $\omega > 0$ case: focus on $h > 0$ regimes for $1/3 > \omega > 0$ on (a) panel, focus on $h > 0$ regimes for
$\omega > 1/3$ on (b) panel and focus on $h < 0$ regimes for $\omega > 1/3$ on (c) panel.
 Dashed blue line corresponds to the $\rho_0$ from (\ref{D2_rho0}).
Dark blue area corresponds to
the unphysical $\rho < 0$ initial conditions. Green curve points location of $\dot h = 0$ while black -- $\dot H = 0$
(see the text for more details).}\label{D2_6}
\end{figure}

They are presented in Figs.~\ref{D2_5}--\ref{D2_6}. In Fig.~\ref{D2_5} we presented $\omega < 0$ case with focus on $h > 0$ regimes on (a) panel and on $h < 0$ regimes on (b) panel;
we also presented central area between $P_{iso}$ at $H=0$, $h=0$ and $E_{iso}$ on panel (c). Similar to the previous figures, dark blue area corresponds to
the unphysical $\rho < 0$ initial conditions; green curve shows the location of $\dot h = 0$ while black -- of $\dot H = 0$; similar to Fig.~\ref{D2_4}, dashed blue line points
to the location of $\rho_0$ from (\ref{D2_rho0}). From Figs.~\ref{D2_5}(a, b) one can clearly see all three possible past asymptotes -- $P_{(0, 1)}$ at $H > 0$, $h\to +\infty$
(and at $H < 0$, $h\to -\infty$), $P_{(1, 0)}$ at $H\to +\infty$, $h\to 0$ (and at $H\to -\infty$, $h\to 0$) and $K_3$. The possible future asymptotes could be seen from all panels of
 Fig.~\ref{D2_5} -- they are isotropic exponential solution $E_{iso}$ and isotropic power-law expansion $P_{iso}$. The separation of different regimes over ($H, h$) plane could be
 seen in Fig.~\ref{D2_5}(c) -- they are $P_{(0, 1)} \to E_{iso}$, $P_{(0, 1)} \to P_{iso}$ (coming from $h\to +\infty$ on $H > 0$) and $P_{(1, 0)} \to E_{iso}$, $K_3 \to P_{iso}$ (coming
 from $H\to +\infty$).

 One can note in Fig.~\ref{D2_5}(c), that apart from two described crossings of $\dot h = 0$ (green curve) with $\dot H = 0$ (black curve) -- $P_{iso}$ and $E_{iso}$ -- there is one
 more more inbetween. As we mentioned earlier, these crossing give rise to exponential solutions -- $P_{iso}$, located at ($0, 0$) is trivial solution while isotropic exponential
 solution is nontrivial one. So that formally the third crossing also could be obtained under certain manipulation (it is located at $\{h\equiv H = \pm 1/\sqrt{-12\alpha}\}$), but
 it is not exponential solution in the general sense. One can also note that it happening at $\rho = \rho_0$ which makes it unstable, so that it closer to CVS from $\alpha > 0$ then
 to ``normal'' exponential solutions.

 The situation with $\alpha < 0$, $\omega > 0$ is not much different from the described above $\omega < 0$. We presented the corresponding phase portraits in Fig.~\ref{D2_6}:
 $1/3 > \omega > 0$ case with focus on $h > 0$ we presented on (a) panel, $\omega > 1/3$ case with focus on $h > 0$ we presented on (b) panel and $\omega > 1/3$ case with focus on
 $h < 0$ we presented on (c) panel. Comparing Figs.~\ref{D2_6}(a, b) with Fig.~\ref{D2_5}(a) shows us little difference and these differences do not change the regimes; the
 same could be stated about comparison of Fig.~\ref{D2_6}(c) with Fig.~\ref{D2_5}(b); the situation in the central region -- between $P_{iso}$ and $E_{iso}$ -- presented
 in Fig.~\ref{D2_5}(b), remains the same in $\omega > 0$ cases as well, including the regimes. The only big difference lies in the presence of CVS in Fig.~\ref{D2_6}(c) -- but, as it
 happening in $\rho < 0$ area (as we found above), it has no impact on the physical regimes.

 This concludes our study on the regimes in $D=2$ case with matter in form of perfect fluid. We have found a number of non-singular regimes: for $\alpha > 0$ they are
 $P_{(1, 0)} \to P_{iso}$  and $K_3 \to E_{3+2}$ for $\omega < 1/3$ and $P_{iso} \to P_{iso}$ for $\omega > 1/3$. For $\alpha < 0$ all regimes are non-singular and they are the same
 for both $\omega > 1/3$ and $\omega < 1/3$: $P_{(0, 1)} \to E_{iso}$, $P_{(0, 1)} \to P_{iso}$, $P_{(1, 0)} \to E_{iso}$ and $K_3 \to P_{iso}$. One can also see that only one of them,
 $K_3 \to E_{3+2}$, has
 realistic compactification. It exists for $\alpha > 0$, $\omega < 1/3$ and the measure of the initial conditions leading to this regime is increasing with growth of $\omega$,
 reaching its maximum at $\omega = 1/3 - 0$.

\section{Discussions}
\label{dis}

Let us summarize the results obtained through the paper. We have considered $D=1, 2$ (the number of extra dimensions) flat cosmological models with the spatial section to be a product
of three- and extra-dimensional subspaces. As a source we consider perfect fluid, which exist in the entire space (not only in the three-dimensional subspace).
We derived analytically the locations of the exponential solutions and nonstandard singularities and plot phase portraits of the
evolutionary trajectories to find all possible regimes for the entire range of initial conditions and parameters.

The results for $D=1$ demonstrate that for $\alpha > 0$ the non-singular regimes are $P_{(1, 0)} \to P_{iso}$, $P_{(0, 1)} \to P_{iso}$ and
$P_{iso} \to P_{iso}$, but neither of them have realistic compactification; there are no nonstandard singularities for $\alpha > 0$. For $\alpha < 0$ situation changes -- there are
nonstandard singularities and so singular regimes emerge. Nonsingular regimes for $\alpha < 0$ include $P_{(0, 1)} \to P_{iso}$ and $P_{(1, 0)} \to E_{iso}$ and both of them exist only
for $\omega < 1/3$; for $\omega > 1/3$ all regimes are singular (have nonstandard singularity as either past or future asymptote or both). But despite the fact that there is a number
of non-singular regimes, neither of them has realistic compactification, making $D=1$ case degenerative in this sense.

The results for $D=2$ somewhat opposite to $D=1$: for instance, for $D=1$ nonstandard singularities exist only for $\alpha < 0$ while for $D=2$ they exist only for $\alpha > 0$. Also,
in $D=1$ case nonstandard singularities are located on a fixed positions (straight lines) while in $D=2$ they located along nontrivial curves. The $\alpha > 0$ cases has a lot of
singular regimes, but there are three nonsingular: $P_{(1, 0)} \to P_{iso}$  and $K_3 \to E_{3+2}$ for $\omega < 1/3$ and $P_{iso} \to P_{iso}$ for $\omega > 1/3$. Among them
$K_3 \to E_{3+2}$ has realistic compactification -- the only regime with realistic compactification discovered through the course of this paper. For $\alpha < 0$ all the regimes are
nonsingular and they are the same for both $\omega > 1/3$ and $\omega < 1/3$: $P_{(0, 1)} \to E_{iso}$, $P_{(0, 1)} \to P_{iso}$, $P_{(1, 0)} \to E_{iso}$ and $K_3 \to P_{iso}$; one can see
that neither of them has realistic compactification.

In the course of the investigation we have faced a number of observations and issues which require additional discussion and we are going to do it now. First of all, one can note that
in $D=1$ and $\alpha > 0$, $D=2$ there is a change of regimes while we cross $\omega = 1/3$, and mainly this involve past asymptote regimes. The reason behind it was several times
mentioned in the course of the paper and now we want to explain it in detail. In~\cite{grg10} we have demonstrated that $\omega = 1/3$ is a critical value for the equation of
state which separate two dynamically distinct cases in (pure) GB cosmology in the presence of the matter in form of perfect fluid -- the cases of GB dominance and matter dominance.
In particular, for $\omega < 1/3$ the past asymptote (high-energy) is GB-dominated while the future asymptote (low-energy) changes to matter-dominated. For $\omega > 1/3$ the
situation is opposite -- the past asymptote (high-energy) is matter-dominated while the future asymptote (low-energy) is GB-dominated. In~\cite{grg10} we clearly demonstrated that
for $\omega > 1/3$ initial singularity is isotropic one ($P_{iso}$) while for $\omega < 1/3$ it is standard $K_3$. The same way, for $\omega > 1/3$ late-time regime is $K_3$
while for $\omega < 1/3$ is isotropic power-law regime $P_{iso}$. But this situation cannot be directly applied to our case -- we have EGB gravity so that the low-energy regime
is governed by GR -- in vacuum the regime should be $K_1$ but in the presence of matter it is $P_{iso}$. And these two $P_{iso}$ are actually different -- the former of the
mentioned is $P_{iso}^{GB}$ and it is governed by GB in the presence of matter while the latter is $P_{iso}^{GR}$ -- isotropic power-law expansion governed by GR in the presence
of matter. Also, for GR, as we mentioned in~\cite{grg10}, the critical equation of state in $\omega = 1$ which corresponds to Jacobs solution~\cite{Jacobs}, and since we
do not consider
$\omega > 1$, for all $\omega$ under consideration the late-time asymptote (if nonsingular!) is $P_{iso}^{GR}$. From this explanation one can understand why only past
(high-energy) asymptote is affected by the change of $\omega \gtrless 1/3$. And well according to the results from~\cite{grg10}, for $\omega > 1/3$ the past asymptote becomes $P_{iso}$
in $D=1$ and $\alpha > 0$, $D=2$ cases. However, for $\alpha < 0$, $D=2$ it is not the same and the next observation clarify the reason behind it.

The mentioned $\alpha < 0$, $D=2$ case has another interesting feature -- the maximal value for the density for a open range of the initial conditions ($H^2 > -1/(36\alpha)$).
In all
the other cases -- entire $D=1$ as well as $\alpha > 0$, $D=2$ -- the value for the density is unbounded. In a sense, the situation is similar to the $\Lambda$-term
cases~\cite{my16b, my17a, my18a}, when the solution does not exist in a certain region of ($H, \Lambda$) for a given $\alpha$. The reason both in the $\Lambda$-term cases and
here is the
same -- certain quadratic equation has negative discriminant in this certain region. But in our case, as now we have more degrees of freedom, the situation is more complicated,
and so the analogue is not direct.

Still, the maximal density exist for all $H^2 > -1/(36\alpha)$, reaching $\rho_0 \to \infty$ at $H\to\pm\infty$, so that formally at $H\to\infty$ the maximal density limit is
lifted. But in reality it is not -- indeed, any small violation from $H=\infty$ gives rise to $\rho_0 < \infty$ and so the matter-dominated regime would be destroyed. That is the
reason we do not see it -- due to limit on the maximal density it cannot be reached in EGB gravity. Still, in pure GB it formally exists (see~\cite{grg10}).

In the course of study we detected what we call constant-volume solution (CVS) -- anisotropic exponential regime whose dimensions are expanding in a way so that the volume of
the space is kept constant. In~\cite{CST2} we investigated this regime in great detail for the totally anisotropic flat (Bianchi-I-type) metrics. In the course of study we
encountered CVS in $\Lambda$-term EGB case (see~\cite{my16b, my17a, my18a}), and there this regime is accessible (so that it could be initial or final asymptote for a trajectory)
and has directional stability (see~\cite{my17a} for details).
Unlikely the $\Lambda$-term case, in our current investigation CVS is inaccessible -- it is a pole-like singular point on the phase space (see e.g. Fig.~\ref{D1_3}(c));
though, the solution itself formally exist and we derived it for each particular case. The reason behind this difference lies in different number of independent variables between
vacuum and $\Lambda$-term cases (one independent variable) on one hand and the case with matter in form of perfect fluid (two independent variables) on the other hand. It seems
that this  additional bound in vacuum and $\Lambda$-term cases plays the crucial role, but formal in-depth investigation of this feature is required and we are going to perform
it shortly.

Finally we can compare our results with those obtained in EGB vacuum and $\Lambda$-term cases. Indeed, formally assuming $\rho \to 0$ we should obtain the regimes from the vacuum
case, but there could be deviations, considering possible changes as $\omega \gtrless 1/3$. Also, for $\omega \to -1$ (assuming it is not singular!) we expect to
recover $\Lambda$-term regimes, but
only for $\Lambda > 0$, as with $\rho > 0$ we cannot recover AdS regimes. To perform the comparison, we use our previously obtained results for vacuum~\cite{my16a} and
$\Lambda$-term~\cite{my16b, my17a} regimes as well as their revision in~\cite{my18a}.

So in $D=1$ the vacuum regimes are~\cite{my16a, my18a} $P_{(1, 0)} \to K_1$ for $\alpha > 0$ and $P_{(1, 0)} \to E_{iso}$, $nS \to E_{iso}$ and $nS \to K_1$ for $\alpha < 0$,
starting from large $H$.
Our results for $D=1$, $\alpha > 0$ are presented in Fig.~\ref{D1_1} and one can see that $\rho \to 0$ and $\omega < 1/3$ regimes would be $P_{(1, 0)} \to P_{iso}$ -- the same
past asymptote as in the vacuum case but
different future asymptote. Indeed, for exact $\rho \equiv 0$ we would have $K_1$ but even tiny nonzero amount of matter destroy this regime turning it into $P_{iso}$ eventually.
For $\omega > 1/3$ the past asymptote is already matter-dominated so that it is $P_{iso}$ (as already noted above, different from the low-energy $P_{iso}$) -- neither of the
regimes are the same as in the vacuum case.
For $\alpha < 0$ and $\omega < 1/3$ the
situation is the same -- high-energy regimes in our case are the same as in the vacuum case while the low-energy regime is replaced with $P_{iso}$ (see Figs.~\ref{D1_2}
and \ref{D1_3}(a, b)). For $\omega > 1/3$ the situation is also similar to that at $\alpha > 0$ -- neither of the regimes are the same as in the vacuum case. So that the comparison
of the vacuum and perfect fluid regimes for $D=1$ reveals that for $\omega < 1/3$ the past asymptote is the same for both cases while the future asymptote is different; for
$\omega > 1/3$ both asymptotes are different.

The structure of the regimes for $D=1$ $\Lambda$-term case is more complicated~\cite{my16b, my18a}.
For $\alpha > 0$, $\Lambda > 0$ the regimes are $P_{(1, 0)} \to E_{iso}$ and $P_{(0, 1)} \to E_{iso}$;
the corresponding regimes could be derived from Fig.~\ref{D1_1}(a). For $\omega \to -1$ the density would become constant and so $P_{iso}$ could not be obtained and would be replaced with $E_{iso}$. This way, only past asymptotes are the same in both $\Lambda$-term and perfect fluid cases -- similar to what we had previously
from the
comparison with vacuum cases. For $\alpha < 0$, $\Lambda > 0$ the situation is more complicated, as there are two subcases -- $\alpha\Lambda \gtrless -3/2$. For
$\alpha\Lambda < -3/2$ the $\Lambda$-term regimes are $P_{(0, 1)} \to nS$ and $P_{(1, 0)} \to nS$ and they both could be obtained in the perfect fluid case (see Fig.~\ref{D1_2}). For
$\alpha\Lambda > -3/2$ the $\Lambda$-term regimes are $P_{(1, 0)} \to E_{iso}^1$, $nS \to E_{iso}^1$, $nS \to E_{iso}^2$ and $P_{(0, 1)} \to E_{iso}^2$. First two of them could
easily be retrieved (see Fig.~\ref{D1_2}) but not the remaining two. The situation with them resemble the previous case when $P_{iso}$ was replaced with isotropic exponential
solution, so now it is the same but the isotropic exponential regime is different from the used in the first pair. So that for $\alpha\Lambda < -3/2$ both regimes are retrieved
completely while for $\alpha\Lambda > -3/2$ we retrieve all but low-energy regimes -- the same as for $\alpha > 0$.

The cases with $D=2$ have more complicated structure; vacuum regimes~\cite{my16a, my18a} for $\alpha > 0$ are $P_{(1, 0)} \to K_1$ for one and
$K_3 \to E_{3+2} \ot nS \to nS \ot K_1$ for another
branch. Comparing these regimes with those obtained with use of $\rho \to 0$  from Fig.~\ref{D2_1}(c) we can verify that all except the low-energy regime are the same; the
low-energy regime $K_1$ is replaced with $P_{iso}$ for the same reasons as discussed in the $D=1$ case. The regimes in Fig.~\ref{D2_2} are also the same but not in Fig.~\ref{D2_3}, so that we recover all regimes except low-energy one for $\omega < 1/3$ and recover neither for $\omega > 1/3$ -- again, in agreement with $D=1$ results.

The resulting regimes for $\alpha < 0$  $D=2$ vacuum model are $K_3 \to K_1$ for one and \linebreak $P_{(1, 0)} \to E_{iso} \ot nS$ with $K_1 \to nS$ for another branch.
Similarly to $\alpha > 0$
case, some of the regimes could be easily retrieved: $K_3 \to K_1$ is ``transformed'' into $K_3 \to P_{iso}$; $P_{(1, 0)} \to E_{iso}$ is the same in both. Two remaining regimes
are more difficult to link: $K_1$ is replaced with $P_{iso}$, but now there is change of direction -- the vacuum regime $K_1 \to nS$ is replaced with $P_{(0, 1)} \to P_{iso}$ --
as we seen before, $K_1$ is replaced with $P_{iso}$, but here the direction of evolution also has been changed; $nS$ is replaced with $P_{(0, 1)}$. The same change happened with
the remaining regime: $nS \to E_{iso}$ is replaced with $P_{(0, 1)} \to E_{iso}$. So that, part of $D=2$ vacuum regimes also could be obtained from perfect fluid approach, but
with some changes, like mentioned $K_1 \to nS$ regime.

Finally let us connect $D=2$ $\Lambda$-term regimes~\cite{my16b, my18a} with those obtained from the perfect fluid approach in the current paper. For $\Lambda > 0$, $\alpha > 0$
regimes for $D=2$ have ``fine structure''~\cite{my16b} -- different sequence of regimes depending on $\alpha\Lambda \in [15/32; 1/2]$. We generally do not recover this fine
 structure; and the recovered regimes along $\rho = \const \equiv \Lambda$ do not correspond to any particular $\alpha\Lambda$ from the $\Lambda$-term case. We can see the same
 two nonstandard singularities in both cases, but no exponential solutions in the perfect fluid case. Formally one could retrieve them , but their emergence would be quite
 nontrivial and cannot be described within perfect fluid approach.
 This could indicate the fundamental difference between $\Lambda$-term and perfect fluid cases.

 The regimes for $\Lambda > 0$, $\alpha < 0$ are different for $\alpha\Lambda \gtrless -5/6$: for $\alpha\Lambda \geqslant -5/6$ there are two (one for exact equality) isotropic
 exponential solutions as future asymptotes while for $\alpha\Lambda < -5/6$ there are anisotropic exponential solutions. The past asymptotes $P_{(1, 0)}$ and $K_3$ are
 recovered perfectly while
 $nS$ are replaced with $P_{(0, 1)}$ -- similar to $\alpha > 0$ case. Of the future asymptotes, similar to the previous $\alpha > 0$ $D=2$ $\Lambda$-term case, anisotropic
  exponential solutions cannot be recovered -- for the same reasons. On the contrary, isotropic exponential solutions could, with the similar technics as in $D=1$ $\Lambda$-term
  case (see above).

  To conclude, vacuum regimes could be reconstructed from $\omega < 1/3$ perfect fluid regimes with some changes in the future asymptotes; $\omega > 1/3$ regimes have no
  connections with vacuum regimes. $\Lambda$-term regimes could be reconstructed only partially for $D=1$ case; $D=2$ case regimes are even less possible to reconstruct, in
  particular, $\Lambda$-term anisotropic exponential solutions could not be obtained from the perfect fluid approach.

\section{Conclusions}
\label{conc}

In this paper we have described the regimes which emerge in $D=1, 2$ (the number of extra dimensions) flat cosmological models in EGB gravity with the matter source in form of
perfect fluid. The results of the investigation
suggest that in $D=1$ there are no regimes with realistic compactification while in $D=2$ there is. It is the transition from high-energy Kasner regime $K_3$ to exponential regime with
expansion of three- and contraction of the remaining two-dimensional subspaces. It is very remarkable fact, as it could lead to a possible explanation of both the Dark Energy problem
and features from effective string theory in the Early Universe within the same theory. Truth be told, we observed the late-time acceleration in both $\Lambda$-term and vacuum models; and if
in the former of them accelerated expansion is not something amazing, in the latter it is, as in the GR accelerated expansion without any matter source is absent. Nevertheless,
neither of these two could hardly describe the current state our Universe -- we definitely observe existence of the ordinary matter. That is why the results of our current paper
are important
-- we  demonstrated that (at least in $D=2$) the model with matter in form of perfect fluid and without $\Lambda$-term can have accelerated expansion phase at late times.

We shall proceed with the investigation of the same model in EGB but it higher dimensions, as well as to consider higher Lovelock contributions -- this would give us answer
on the question if this behavior is common and how much is it spread with respect to the parameters and initial conditions.

We have noted an interesting feature -- in $\alpha < 0$, $D=2$ case there exists maximal density for the matter, and due to this fact for $\omega > 1/3$ the past asymptote is
GB-dominated, opposite to all other cases, where for $\omega > 1/3$ the past asymptote is matter-dominated. In higher-dimensional EGB cosmologies the structure of the branches
would be even more complicated so that the dynamics could be even more interesting and there could be other unexpected features, which makes this case more interesting to consider.

\end{document}